\documentclass[12pt,letterpaper]{article}

\usepackage{amsmath,amssymb,array,calc,rotating,epsfig,psfrag}
\usepackage[nosort]{cite}

\newcommand\be{\begin{equation}}
\newcommand\ee{\end{equation}}
\newcommand{\bea}{\begin{eqnarray}}
\newcommand{\eea}{\end{eqnarray}}

\newcommand{\nn}{\nonumber}
\newcommand{\pd}{\partial}
\newcommand{\ul}{\underline}
\def\id{\protect{{1 \kern-.28em {\rm l}}}}

\def\cV{{\cal V}}
\def\ve{\varepsilon}

\font\mybb=msbm10 at 12pt

\def\bb#1{\hbox{\mybb#1}}

\def\id{\protect{{1 \kern-.28em {\rm l}}}}

\begin{document}

\begin{titlepage}
\begin{center}
\hfill MCTP-05-82  \\
\hfill {\tt hep-th/0506191}\\
\vskip 15mm

{\Large {\bf $R^4$ Corrections to Heterotic M-theory \\[3mm] }}

\vskip 10mm

{\bf Lilia Anguelova, Diana Vaman }

\vskip 4mm
{\em Michigan Center for Theoretical Physics}\\ 
{\em Randall Laboratory of Physics,
University of Michigan}\\ {\em Ann Arbor, MI 48109-1120, USA} \\
{\tt anguelov@umich.edu, dvaman@umich.edu}\\

\vskip 6mm

\end{center}

\vskip .1in

\begin{center} {\bf ABSTRACT }\end{center}

\begin{quotation}\noindent

We study $R^4$ corrections in heterotic M-theory. We derive to order $\kappa^{4/3}$ the induced modification to the K\"{a}hler potential of the universal moduli and its implications for the soft supersymmetry breaking terms. The soft scalar field masses still remain small for breaking in the $T$-modulus direction. We investigate the deformations of the background geometry due to the $R^4$ term. The warp-factor deformation of the background $M_4\times CY(3)\times S^1/Z_2$ can no longer be integrated to a fully non-linear solution, unlike when neglecting higher derivative corrections. We find explicit solutions to order $\kappa^{4/3}$ and, in particular, find the expected shift of the Calabi-Yau volume by a constant proportional to the Euler number. We also study the effect induced by the $R^4$ terms on the de Sitter vacua found previously by balancing two non-perturbative contributions to the superpotential, namely open membrane instantons and gaugino condensation. 
To order $\kappa^{4/3}$ all induced corrections are proportional to the Euler 
number of the Calabi-Yau three-fold.

\end{quotation}
\vfill

\end{titlepage}

\eject

\tableofcontents

\section{Introduction}

Heterotic strings have long provided the most promising  candidate for unified description of phenomenology despite persisting problems. It was realized in \cite{EW,BD} that some of these problems can be resolved if one considers their strongly coupled limit, given by M-theory on an interval \cite{HW}. The corresponding four-dimensional compactification \cite{LOW}, called heterotic M-theory, has received a great deal of attention. Especially interesting, in view of the astronomical observations indicating a positive cosmological constant and an exponential expansion of the early universe, are the recently found de Sitter \cite{BCK} and assisted inflation \cite{BBK} solutions. Heterotic M-theory has the very distinctive feature that unlike the weakly coupled case it does not allow vanishing background flux. 

In string theory, nonzero fluxes play a significant role in the resolution of the moduli stabilization problem. The latter occurs in purely geometric compactifications due to the lack of a potential for the many scalar fields that originate from deformations of the internal Calabi-Yau  manifold. These moduli are of two types depending on whether they parametrize the complex structure or the K\"{a}hler structure deformations. Stabilizing them is essential for predictability of the four-dimensional coupling constants and also for avoiding decompactification of the internal space.
It was realized in the context of type IIB \cite{GKP} that background fluxes generically lift the flat directions of the complex structure moduli by generating a superpotential for them. But this superpotential does not depend on the K\"{a}hler moduli. So in order to stabilize the latter one has to resort to quantum corrections.\footnote{In type IIA, however, all moduli can be stabilized classically \cite{DGKT}.} There are two kinds of nonperturbative effects that can create a potential for the Kahler moduli: D-brane instantons and gaugino condensation. It was argued in \cite{KKLT} that using these and nonzero NS-NS and RR fluxes one can fix all moduli. 

Another modification of the K\"{a}hler potential is due to $\alpha'$ corrections \cite{BBHL}, which appear as higher derivative terms in the string effective action. Typically though, their contribution was expected to be suppressed in the large volume limit. However, this was shown to be too naive in \cite{BB}. That work argued that, as classically the K\"{a}hler moduli are flat directions of the potential, the perturbative, $\alpha'$, corrections are generically the leading ones even at large volume. Since they dominate the non-perturbative contributions, their presence alters qualitatively the structure of the scalar potential.

In M-theory there are also higher derivative corrections to the eleven-dimensional supergravity action.\footnote{They can be deduced from duality with ten-dimensional string theory \cite{GV,DLM} or from superparticle scattering amplitudes in $d=11$ \cite{suppart}.} When compactified on $\bb{R}^{1,9}\times \bb{S}^1\!/\bb{Z}_2$, the theory has an expansion in powers of $\kappa^{2/3}$ \cite{HW} with $\kappa$ being the gravitational coupling constant. Completing the effective action at orders $\kappa^{4/3}$ and higher encounters technical problems that may be overcome in the approach of \cite{IM}. Our main focus in this work is the eleven-dimensional $R^4$ term which appears at ${\cal O}(\kappa^{4/3})$. Naturally, it is expected to correct the K\"{a}hler potential for the moduli fields of the appropriate Calabi-Yau three-fold compactification to four dimensions. We compute this correction as well as its implications for the soft supersymmetry breaking terms in the four-dimensional effective theory. As in \cite{LOW2}, the scalar soft masses still do not receive any tree level contribution for supersymmetry breaking in the $T$-modulus direction.

The $R^4$ term should also affect the geometry of the background solution. To first order in $\kappa^{2/3}$, the solution of heterotic M-theory was studied in \cite{EW,LOW}. Due to the $E_8$ gauge fields propagating on the two boundaries, the Bianchi identity of the supergravity three-form is modified, leading to nonzero background flux and a generically non-K\"{a}hler deformation of the initial $CY(3)$. Clearly, this is the strong-coupling description of the weakly coupled heterotic string compactifications with torsion \cite{torsion}.\footnote{Recent improvement in their understanding is due to \cite{Lust1}, where it was shown that up to order ${\cal O}(\alpha'^2)$ the supersymmetry conditions and Bianchi identities imply the field equations (recall that generically this is true only for maximally supersymmetric backgrounds), and also \cite{Lust2}, where the role of gaugino condensation for the effective four-dimensional superpotential was clarified.} Progress in the explicit construction of the latter non-K\"{a}hler backgrounds was achieved only very recently \cite{nonK}.\footnote{Strictly speaking, the work of \cite{nonK} gives backgrounds only for the $SO(32)$ heterotic string. But presumably one can use similar methods for the $E_8\times E_8$ case. We thank Radu Tatar for a discussion on that issue.} On the other hand, not much is known about the explicit form of their strongly coupled lift at orders higher than $\kappa^{2/3}$. One can simplify things by considering only warp factor deformations of the eleven-dimensional metric, which in particular means only conformal deformations of the Calabi-Yau.\footnote{Of course, that implies certain restrictions on the three-form flux.} In this case, a non-linear background containing corrections of all orders in $\kappa^{2/3}$ was obtained in \cite{CK}, without taking into account any higher derivative terms in the eleven-dimensional effective action. This solution was used in an essential way in \cite{BCK,BBK}. As the eleven-dimensional $R^4$ term appears already at second order in the $\kappa^{2/3}$ expansion, a valid question is how it modifies the solution of \cite{CK}. We will see that certain relations between warp factors have to be different in the present case. Also, the presence of higher derivative terms opens up the possibility of turning on simultaneously different flux components while still preserving the only-warp-factor character of the geometric deformation, which was not possible before. We find the anticipated shift of the Calabi-Yau volume by a constant proportional to the Euler number, although it is unlikely that this would resolve the singularity of the non-linear background of \cite{CK}, appearing at a point along the interval $\bb{S}^1\!/\bb{Z}_2$ at which the Calabi-Yau volume shrinks to zero \cite{CK2}. 


The non-linear solution of \cite{CK} was used in \cite{BCK} to find de Sitter vacua in heterotic M-theory by balancing gaugino condensation against membrane instantons. One might expect that, similarly to the string theory case \cite{BB}, here too the $R^4$ corrections will be dominant over these non-perturbative effects. However, this does not happen essentially because the no scale structure of the K\"{a}hler potential for the $T$-modulus is preserved by the $R^4$ term. On the other hand, taking into account higher derivative corrections means that one can not integrate the solution to all orders in $\kappa^{2/3}$ but instead should always work only to the appropriate level of accuracy. 
Therefore, it is worth revisiting the considerations of \cite{BCK}, that led to the existence of de Sitter vacua, in the context of the $R^4$ corrected
effective action. 
We will see that de Sitter vacua still exist although they appear to have a 
much smaller cosmological constant as a result of employing the truncated to
${\cal O}(\kappa^{4/3})$ solution.  
The $R^4$ induced corrections are of the order of a small percentage and strengthen/weaken the positive value of the energy density for a positive/negative value of the CY Euler number.

The present paper is organized as follows. In Section 2 we review necessary material about the linear solution of \cite{EW,LOW} and write down the correction to the K\"{a}hler potential due to the $R^4$ term. A detailed derivation of this correction is given in Appendix A. In Section 3 we calculate the $R^4$ induced contributions to the soft supersymmetry breaking masses of the gravitino, gaugino and scalar fields  and also to the trilinear couplings. In section 4 we consider the influence of the $R^4$ correction on the geometric background. In Subsection 4.1 we find solutions for the case of a metric deformation given by warp factors only. In Subsection 4.2 we address generic non-K\"{a}hler deformations of the initial Calabi-Yau. We elaborate more on that in Appendix B, where we derive the appropriate generalized Hitchin flow equations, that constitute the first steps towards finding explicit solutions for the strongly coupled limit of heterotic strings on generic non-K\"{a}hler manifolds. Finally, Section 5 is dedicated to the investigation of the minima of the scalar potential in the presence of the $R^4$ terms.

\section{$R^4$ terms and Horava-Witten theory}
\setcounter{equation}{0}

\bigskip

The strongly coupled limit of the heterotic $E_8 \times E_8$ string theory was argued in \cite{HW} to be given at low energies by eleven-dimensional supergravity on $\bb{R}^{1,9}\times \bb{S}^1\!/\bb{Z}_2$. To obtain an effective field theory on four-dimensional Minkowski space, one further compactifies on a Calabi-Yau three-fold. The eleven-dimensional action has an expansion in powers of $\kappa^{2/3}$, where $\kappa$ is the gravitational coupling constant. More precisely, it is of the form\footnote{It has been argued in \cite{BD,LOW} that the expansion in powers of $\kappa^{2/3}$ is in fact an expansion in the dimensionless quantity $\epsilon=\frac{2\pi L}{3V_v^{2/3}} (\frac{\kappa}{4 \pi})^{\frac{2}{3}}$, where $L$ is the length of the interval and $V_v$ is the Calabi-Yau volume at the visible boundary.} 
\be \label{expansion}
S_{11d} = \frac{1}{\kappa^2} \left( S^{(0)} + \kappa^{2/3} S^{(1)} + \kappa^{4/3} S^{(2)} + ... \right) 
\ee
with $S^{(0)}/\kappa^2$ being the well-known Cremmer-Julia-Scherk action \cite{CJS}:
\be \label{S_0}
S_0=-\frac{1}{2\kappa_{11}^2}\int d^{11}x\bigg(R\wedge *1-\frac 12 G_{(4)}\wedge
*G_{(4)}-\frac{1}{6}C_{(3)}\wedge G_{(4)}\wedge G_{(4)}\bigg) \, ,
\ee
where $G_{(4)} = d C_{(3)}$. The reduction to four dimensions was performed in \cite{LOW} completely to order $\kappa^{2/3}$, meaning the term $S^{(1)}$, and partially at order $\kappa^{4/3}$, meaning that only some of the contributions to $S^{(2)}$ were found. Since this will be important in the following, let us recall the main features of the results of \cite{LOW,EW}.

To start with, the presence of the two boundaries in eleven dimensions modifies the Bianchi identity of the three-form $C_{(3)}$ and hence a vanishing C-field background is not a solution anymore. As a consequence of this, the eleven-dimensional metric acquires nontrivial warp factors. In other words, the initial direct product $M_4\times CY(3) \times \bb{S}^1\!/\bb{Z}_2$ gets deformed to a nontrivial fibration of a (in general non-K\"{a}hler) 6d  manifold over the interval $\bb{S}^1\!/\bb{Z}_2$. To first order the deformed metric has the form \cite{EW}:
\be \label{def_met}
ds^2_{11d} =  g_{IJ}^{(0)} dx^I dx^J + b \eta_{\mu \nu} dx^{\mu} dx^{\nu} + h_{mn} dx^m dx^n + k (dx^{11})^2 \, ,
\ee
where $I,J=1,...,11$; $\mu,\nu=1,...,4$ and $m,n=5,...,10$. The universal moduli\footnote{As in \cite{LOW}, these are the only moduli we will concentrate on.} (meaning those that are independent of the particular $CY(3)$ that one is considering) of the original direct product  metric $g_{IJ}^{(0)}$ are the volume of the Calabi-Yau space and the size of the interval (orbifold) $\bb{S}^1\!/\bb{Z}_2$. Hence one can write
\be \label{g0}
g_{IJ}^{(0)} dx^I dx^J = g_{\mu\nu}dx^\mu dx^\nu + e^{2a} \tilde{g}_{mn} dx^m dx^n +e^{2c} (dx^{11})^2 \, ,
\ee
where the Calabi-Yau volume and the interval size are parametrized
in terms of the scalar fields $a(x^\mu), c(x^\mu)$. The deformed metric (\ref{def_met}) has the same moduli, but the dependence on them is more complicated since generically:
\be
b = b(a,c) \, , \qquad h_{mn} = h_{mn} (a,c) \, , \qquad k = k (a,c) \, .
\ee
We will not need the explicit form of the first order solution (\ref{def_met}), found in \cite{LOW}. We only note that it is such that {\it there are no} order ${\cal O} (\kappa^{2/3})$ corrections to the effective action that are due to the deformed metric or $C$-field \cite{LOW}. As shown in \cite{LOW}, the only contribution at that order comes from the gauge multiplets propagating on the ten-dimensional boundaries. The contribution of the latter at order $\kappa^{4/3}$ is the  ${\cal O} (\kappa^{4/3})$ part of the action that was found in \cite{LOW}. To complete the effective action at this order one has to take into account two more contributions: from higher derivative terms and from higher order deformations of the background metric. The first will be our immediate concern; the second we will address in Section 4.

The lowest order higher derivative corrections to the eleven-dimensional supergravity action are the $R^4$ term \cite{GV} and its superpartner $C \wedge R^4$ \cite{DLM}:\footnote{Actually, in Horava-Witten theory there is another contribution: Gauss-Bonnet $R^2$ terms localized on the two ten-dimensional boundaries \cite{LOW}. However, they give rise to higher (than two) derivative terms in the four-dimensional effective action and similarly to \cite{LOW} we omit such contributions in the following.}
\be \label{S1}
S_1=-b_1 T_2\int d^{11}x \sqrt{-g}
\bigg[t_8\cdot t_8 RRRR-\frac 14 E_8-4\epsilon_{11}C_{(3)}
\!\!\left(tr R^4-\frac 14(tr R^2)^2\right)\!\bigg],
\ee
where 
\be
E_8=\frac 1{3!}\epsilon_{I J K M_1\dots M_8} \,
\epsilon^{I J K N_1 \dots N_8}{R^{M_1 M_2}}_{N_1 N_2}
{R^{M_3 M_4}}_{N3 N_4} {R^{M_5 M_6}}_{N_5 N_6} {R^{M_7 M_8}}_{N_7 N_8} \, .
\ee
The constants and parameters in the action $S_1$ are defined by\footnote{We use the notation and conventions of \cite{AT}.}
\be
2\kappa_{11}^2=(2\pi)^5 l_{11}^9, \qquad \!\!l_{11}=(2\pi g_s)^{1/3}{\alpha'}^{1/2},\qquad \!\!T_2=\frac{1}{2\pi l_{11}^3},\qquad \!\!b_1=\frac{1}{(2\pi)^4 3^2 2^{13}}.
\ee
For future convenience we also introduce the notation $\ve \equiv b_1 T_2 \kappa^2$. Clearly $T_2 = (2 \pi)^{2/3} (2 \kappa_{11}^2)^{-1/3}$ and hence $S_1$ appears at order $\kappa^{4/3}$ in the expansion (\ref{expansion}). In the following, our goal will be to extract its contribution to the moduli space metric of the four-dimensional effective theory.

So far we have introduced only half of the universal moduli, namely $a$ and 
$c$. The other two scalar fields, $\sigma_S$ and $\sigma_T$, are axions that  
arise from the eleven-dimensional 3-form $C$ via the identifications
\be \label{s_ph}
C^{(0)}_{mn 11} = \frac{1}{6\sqrt{2}} \sigma_T \,J_{mn} \, , 
\qquad \sqrt{2} *_4 dB = d\sigma_S \qquad {\rm with} \qquad B_{\mu \nu} = 6 C^{(0)}_{\mu \nu 11} \, ,
\ee
where $J_{mn}$ is the K\"{a}hler form of the $CY(3)$. In the four-dimensional $N=1$ theory the 4 universal moduli make up the bosonic components of two chiral superfields:
\be
S = e^{6a} + i \sigma_S \, , 
\qquad T = e^{\hat{c}} + i \sigma_T\, ,
\ee
where we have redefined $\hat{c} = c + 2a$ for later purposes. 
Due to the superfield structure and the independence of the K\"{a}hler potential $K$ on $ImS$ and $ImT$ (see for example \cite{CKM}), in order to deduce the correction to $K(S,T)$ it is sufficient to only keep track of the kinetic terms for $a,c$. Therefore from now on we ignore the 
$C\wedge R^4$ term and concentrate on the dimensional reduction of the remaining terms in the action (\ref{S1}).

Using that $t_8 t_8 R^4 + 1/4 E_8$ can be recast up to Ricci terms in 
the form $64 (12 Z - RS + 12 R_{IJ} S^{IJ})$ \cite{FPSS}, we rewrite the 
part of the action that we want to reduce as
\be \label{S_1}
S_1=- b_1 T_2\int d^{11}x\sqrt{-g}\bigg[2^6 (12 Z-RS+12R_{IJ} S^{IJ} ) - \frac 12 E_8\bigg] +\dots \, ,
\ee
where
\bea
&&Z_{IJ}=R_{IKLR} R_{JMN}{}^R \bigg(R^{K}{}_P{}^M_{}Q R^{NPLQ}-\frac 12
R^{KN}{}_{PQ}R^{MLPQ}\bigg),\nonumber\\
&&S_{IJ}=-2R_{I}{}^{MKL}R_J{}^P{}_{K}{}^QR_{LPMQ}+
\frac12R_I{}^{MKL}R_{JMPQ}R_{KL}{}^{PQ}\nonumber\\
&&~~~~~~~~-R_I{}^K{}_J{}^LR_{KMNQ}R_L{}^{MNQ} \, , \nn \\
&&~~Z=Z_{IJ} g^{IJ} \, , \qquad S=S_{IJ} g^{IJ} \, .
\eea
We will also use that, up to Ricci terms, $S=12(2\pi)^3 Q$  with
the Euler density defined as 
\be
\int_X d^6x \sqrt{det(g_{IJ})} Q = \chi \label{chi} \, ,
\ee 
where $\chi$ is the Euler number of the Calabi-Yau three-fold $X$.

Since the action (\ref{S_1}) is already of order $\kappa^{4/3}$ and going to higher orders in the $\kappa$ expansion of the eleven-dimensional action is beyond the scope of this paper, clearly we have to reduce (\ref{S_1}) on the zeroth order solution for the eleven-dimensional metric, namely the one given in (\ref{g0}). To obtain canonical Einstein term for the four-dimensional action, we have to rescale the metric:
\be
g_{\mu \nu} = e^{-6a-c} \bar{g}_{\mu \nu} \, .
\ee
The details of the reduction are given in Appendix A. Here we only record the final result for the $R^4$ induced correction to the K\"{a}hler potential:
\be
K = -3 \ln (T + \bar{T}) - \ln (S + \bar{S}) - \frac{b_2 \kappa^{4/3} \chi}{6 (S + \bar{S})} \, ,
\ee
where $b_2$ is a numerical coefficient given in (\ref{b2}).
\section{Soft supersymmetry breaking terms}
\setcounter{equation}{0}

The low energy effective action of a four-dimensional theory with $N=1$ supersymmetry is determined by the K\"{a}hler potential $\hat{K}$, the superpotential $W$ and the gauge kinetic function $f$. For the case of the above compactification of the strongly coupled heterotic string, these functions can be read off from the action of \cite{LOW}, which is complete to order $\kappa^{2/3}$ but contains only some of the ${\cal O}(\kappa^{4/3})$ contributions. Expanding to second order in the charged matter fields $C^p$, one has \cite{LOW2}:
\be \label{K_h}
\hat{K} = \kappa_4^{-2} K (S, T, \bar{S}, \bar{T}) + \,Z_{p \bar{q}} (S, T, \bar{S}, \bar{T}) C^p \bar{C}^{\bar{q}} \, ,
\ee
where
\be
K = -\ln (S + \bar{S}) -3\ln(T + \bar{T}) \, , \qquad Z_{p \bar{q}} = \left( \frac{3}{T+\bar{T}} + \frac{\beta}{S + \bar{S}} \right) \delta_{p \bar{q}} \, ,
\ee
and
\be \label{sup_W}
W = \frac{1}{3} \tilde{Y} d_{pqr} C^p C^q C^r + W_{\rm{non-pert.}} \, , \qquad f = S + \beta T \, .
\ee
The four-dimensional gravitational constant is $\kappa_4^2=\kappa_{11}^2/Vl$ and $\beta \sim {\cal O}(\kappa^{2/3})$. The superfields $S$ and $T$ were already introduced in the previous section.\footnote{Abusing notation, we denote both a superfield and its bosonic component with the same letter.} The remaining ones, $C^p$, are chiral superfields that arise from the ten-dimensional $E_8$ gauge fields and are charged under the unbroken gauge group in the visible sector. For phenomenological purposes, $C^p$ are often taken to transform in the {\bf 27} of $E_6$.

Since the action of \cite{LOW} is incomplete at order $\kappa^{4/3}$ so are the above formulae (\ref{K_h})-(\ref{sup_W}). As we saw in Section 2, the additional contribution, coming from the higher derivative $R^4$ correction to the eleven-dimensional effective action, is:
\be \label{del_K}
\delta K = - \frac{b_2 \kappa^{4/3} \chi}{6 (S + \bar{S})} \, .
\ee
This correction has further implications for the soft supersymmetry breaking terms of the four-dimensional theory. Leaving aside for the moment the issue of finding (meta)stable minima of the potential, one can parametrize the supersymmetry breaking by the auxiliary components of the superfields $S$ and $T$, denoted respectively by $F^S$,$F^T$. The general form of the soft terms arising from spontaneous supersymmetry breaking due to non-perturbative effects in the hidden sector was derived in \cite{SWGM}. In a K\"{a}hler covariant language, the tree level formulae for the masses of the gravitino, gaugino, scalar fields and Yukawa couplings are \cite{KL}:
\bea\label{susybr}
&&m_{3/2}^2 = \frac{1}{3} K_{i \bar{j}} F^i F^{\bar{j}} \, , \qquad \qquad \hspace{1cm} m_{1/2} = \frac{1}{2} F^i \pd_i \ln ({\rm Re} f) \, , \nn \\
&&\hspace*{0.15cm}m_{p \bar{q}}^2 = m_{3/2}^2 Z_{p \bar{q}} - F^i F^{\bar{j}} R_{i \bar{j} p \bar{q}} \, , \qquad
A_{pqr} = F^i D_i Y_{pqr} \, ,
\eea
where $i,j$ run over the fields $S,T$ and
\bea
R_{i \bar{j} p \bar{q}} &=& \pd_i \pd_{\bar{j}} Z_{p \bar{q}} - \Gamma_{i p}^s Z_{s \bar{t}} \Gamma^{\bar{t}}_{\bar{j} \bar{q}} \, , \nn \\
D_i Y_{pqr} &=& \pd_i Y_{pqr} + \frac{1}{2} K_i Y_{pqr} - \Gamma_{i(p}^s Y_{qr)s}
\eea
with
\be
Y_{pqr} = e^{K/2} \tilde{Y} d_{pqr} \, , \qquad \Gamma_{i p}^s = Z^{s \bar{t}} \pd_i Z_{\bar{t} p} \, .
\ee

The corrections due to the terms proportional to $\beta$ in $Z_{p \bar{q}}$ and $f$ were computed in \cite{LOW2}. The new contributions coming from the $R^4$ induced change in the K\"{a}hler potential (\ref{del_K}) are:
\bea\label{soft}
\delta m_{3/2}^2 &=& - \frac{\tilde{b} \,|F^S|^2}{9 (S + \bar{S})^3} \, , \qquad \delta m_{1/2} = 0 \, , \nn \\
\delta m_{p \bar{q}}^2 &=& - \frac{\tilde{b} \,|F^S|^2}{3(S + \bar{S})^3 (T + \bar{T})} \delta_{p \bar{q}} \, , \nn \\
\delta A_{pqr} &=& \tilde{b} \left( \frac{F^S}{4 (S + \bar{S})^2} - \frac{F^T}{2(S + \bar{S})(T + \bar{T})} \right) Y_{pqr} \, ,
\eea
where $\tilde{b} = b_2 \kappa^{4/3} \chi$. It may seem surprising that we have obtained corrections of order $\kappa^{4/3}$, whereas those of \cite{LOW2} were of order $\kappa^{2/3}$, since in both cases one starts with an ${\cal O}(\kappa^{4/3})$ correction to the eleven-dimensional action. The resolution of this seeming puzzle is in a rescaling of the matter fields $C^p$, involving a power of $\kappa$ \cite{LOW}, that was needed for the proper normalization of the chiral superfields.

The most significant difference in the patterns of supersymmetry breaking of the weak and strong coupling limits occurs for breaking in the direction of the $T$-modulus, i.e. when $F^T \neq 0$ and $F^S = 0$. In that case, in the weakly coupled heterotic string description only the gravitino mass  acquires a non-vanishing value at tree level. On the other hand, in the strongly coupled heterotic M-theory limit this also happens for $m_{1/2}$ and $A_{pqr}$ \cite{LOW2}. As we see from (\ref{soft}), for $F^S=0$ the $R^4$ term induces an additional contribution to the trilinear couplings $A_{pqr}$, whereas the scalar field masses $m_{p \bar{q}}$ remain small as in the case without higher derivative corrections \cite{LOW2}.

Another observation following from (\ref{soft}) is that the changes we have computed to the soft supersymmetry breaking terms do not spoil universality. Recall that the phenomenological requirement for suppression of flavor changing neutral interactions is satisfied if the masses of the scalar superpartners of the observable fermions (which usually differ from the fermionic components of the charged matter superfields $C^p$ due to various mixing angles) are all almost equal to each other \cite{FCNC}. This condition is achieved for $m^2_{p \bar{q}} \sim \delta_{p \bar{q}}$ and also for $A_{pqr} \sim Y_{pqr}$ with all components $Y_{pqr}$ having the same moduli dependence. Actually, the latter criteria for universality were expected to hold, even without an explicit calculation, due to the general form of the soft terms (\ref{susybr}) and the fact that in the present case $Z_{p \bar{q}}\sim \delta_{p \bar{q}}$. Let us also note that models with more than one K\"{a}hler modulus are generically non-universal and it seems clear that higher derivative corrections would only worsen that effect. An obvious remedy, proposed in \cite{LOW3} and used in many other references, is to consider only Calabi-Yau manifolds with $h^{1,1}=1$.

\section{Deformation of the background geometry}
\setcounter{equation}{0}

In this section we analyze the effect that the higher derivative $R^4$ terms have on the geometric background and $G$-flux of Horava-Witten theory.

Recall that the existence of a solution to linear order in $\kappa^{2/3}$ was shown in \cite{EW} and its explicit form was found in \cite{LOW}. In \cite{CK} a non-linear solution was obtained which contains 
corrections of all orders in $\kappa^{2/3}$. However, none of these works took into account higher derivative corrections to eleven-dimensional supergravity.
Adding the $R^4$ term to the action though, clearly modifies both the 
equations of motion and the supersymmetry variations of the theory. 
The complete supersymmetry transformations are not yet known despite 
significant progress in that direction \cite{PVW}. 
However, the modified gravitino transformation rule was derived on a 
case-by-case basis for compactifications on the following  special holonomy manifolds: 
$CY(3)$ \cite{CFPSS}, $G_2$ \cite{LPST}, $Spin(7)$ \cite{LPST2}.\footnote{
Ref. \cite{DC} also provides a detailed account of the $R^4$-corrected bosonic
 equations of motions in the $Spin(7)$ case.} In each case, the Einstein equations in the presence of the $R^4$ term were recovered from the integrability of the proposed supersymmetry variation by using in an essential way properties particular to the special holonomy manifold under consideration.
Nonetheless, the final result for 
the gravitino variation turned out to be the same for all cases, when 
written in purely Riemannian form, i.e. without the use of any special 
structures. Namely, it was found that the Killing spinor equation is
\be \label{susytr}
\delta \Psi_I = \left( D_I + \ve (\nabla^J R_{IKM_1M_2}) R_{JLM_3M_4} R^{KL}{}_{M_5M_6} \Gamma^{M_1...M_6} \right) \eta = 0 \, ,
\ee
where $\ve$ is a numerical constant times $\alpha'^{\,3}$ in string theory and a numerical constant times $\kappa^{4/3}$ in M-theory.
The derivative $D_I$ is the appropriate extension of the covariant 
derivative $\nabla_I$ with flux terms, i.e. for the case of interest to us:
\be
D_I = \nabla_I + \frac{\sqrt{2}}{288} \left( \Gamma_{IJKLM} - 8 g_{IJ} \Gamma_{KLM} \right) G^{JKLM} \, .
\ee

In the following we will analyze the solutions of (\ref{susytr}). As the order of accuracy to which we work is $\kappa^{4/3}$, we do not need to perform any checks on whether this is also the correct modification for Horava-Witten theory, the reason being that due to $\ve \sim {\cal O}(\kappa^{4/3})$ the three curvature tensors have to be of zeroth order, i.e. they are the curvatures for the direct product $M_4\times CY(3)\times \bb{S}^1/\bb{Z}_2$. In fact, it is most convenient to use the form of (\ref{susytr}) adapted to the Calabi-Yau case:
\be \label{svar}
\delta \Psi_m = (D_m + P_m) \eta \, , \qquad P_m = - \frac{i}{2} 
\ve J_{mn} \pd^nQ \,\Gamma_{11} \, ,
\ee
where $m,n=1,...,6$\, , $Q$ is as before the Euler density in six dimensions, $J_{mn}$ is the K\"{a}hler form of the Calabi-Yau and $\Gamma_{11}$ is the $\Gamma$-matrix in the interval direction. Recall that in terms of the constants in the action (\ref{S1}) $\ve = b_1 T_2 \kappa^2$, where we remind the reader that $T_2 \sim {\cal O}(\kappa^{-2/3})$. Since the $\bb{Z}_2$ projection implies $\Gamma_{11} \eta = \eta$ and also the non-vanishing components of the K\"{a}hler form are $J_{a \bar{b}} = - J_{\bar{b} a} = - i g_{a \bar{b}}$ with $a = 1,2,3$ being the holomorphic indices, we have\footnote{Actually, the positive chirality projection condition $\Gamma_{11} \eta = \eta$ is modified by the warp factor in front of $(dx^{11})^2$ in the full geometry. However, that warp factor is of ${\cal O}(\kappa^{2/3})$ and higher whereas our computation is reliable to ${\cal O}(\kappa^{4/3})$. So in $P_m$ we can simply take $\Gamma_{11} \eta = \eta$.}
\be \label{Pa}
P_a = -\frac{\ve}{2} \pd_a Q \, , \qquad P_{\bar{a}} = \frac{\ve}{2} \pd_{\bar{a}} Q \, .
\ee

Now let us explore the consequences of this additional term to the gravitino variation.

\subsection{Warp factor deformations}

Before addressing general deformations of the initial Calabi-Yau manifold, we will start with the simpler metric ansatz:
\be \label{metr}
ds^2 = e^{b(x^m,x^{11})} \eta_{\mu \nu} dx^{\mu} dx^{\nu} + e^{f(x^m,x^{11})} g_{ln} (x^m) dx^l dx^n + e^{k(x^m, x^{11})} (dx^{11})^2 \, ,
\ee
where the deformation due to fluxes and $R^4$ terms is entirely encoded in three warp factors.

This is precisely the form of the non-linear background of \cite{CK} valid in the absence of $R^4$ corrections. Although in principle we should expand the exponentials and keep terms only up to second order in $\kappa^{2/3}$, we leave (\ref{metr}) for now as it is for easier comparison with the solution of \cite{CK}. 
Generically, the covariantly constant spinor of the full metric, $\eta$, is a deformation of the Calabi-Yau one, $\eta_0$, and we write 
\be
\eta = e^{-\psi(x^m,x^{11})} \eta_0.
\ee 
The Killing spinor equation $\delta \Psi_I = 0$ implies relations between the functions $b, f, k, \psi$ and the $G$-flux components, which are conveniently written in terms of the following quantities\footnote{We note in passing that this parametrization of the four-form flux, first introduced in \cite{EW}, is very reminiscent of the recent G-structure approach to classification of supergravity solutions in various dimensions \cite{G-str,GP}. More precisely, looking at the decomposition of G as $G = - \frac{Q}{3} J\wedge J + J\wedge A + v\wedge (J\wedge W+U)$ in (3.10) of \cite{DP}, one can identify up to numerical coefficients $\alpha$ with $Q$, $\beta_l$ with the 1-form $W$ and $\Theta_{lm}$ with the 2-form $A-\frac{2}{3}QJ+2v\wedge W$ (see eq. (3.12) of \cite{DP}). On the other hand, the 3-form $U$ is given by the components $G_{lmn11}$ such that $G_{lmn11} J^{mn} = 0$. The remaining terms in (3.10) of \cite{DP} vanish due to (3.13) there, which is a consequence of supersymmetry.}:
\be \label{def}
\alpha = G_{lmnp} J^{lm} J^{np} \, , \qquad \beta_l = G_{lmn11} J^{mn} \, , \qquad \Theta_{lm} = G_{lmnp} J^{np} \, .
\ee
It is straightforward to find out how the additional $P_m$ term in (\ref{svar}) modifies the supersymmetry conditions derived in \cite{CK}. The result for the warp factors is:
\bea \label{susyrel}
&&8 \pd_a \psi = -2\pd_ab = \pd_a k = i \frac{\sqrt{2}}{3} e^{-k/2-f} \beta_a = \pd_a f + 2 \ve \pd_a Q \, , \nn \\
&&4 \pd_{11} \psi = - \pd_{11}b = \pd_{11} f = - \frac{\sqrt{2}}{24} e^{k/2-2f} \alpha \, .
\eea
The $G$-flux conditions remain the same as before:
\be \label{G-fl-cond}
\Theta^{\bar{b}}{}_{\bar{a}} = 0 \,\, , \,\, \bar{b} \neq \bar{a} \,\,\, ; \qquad \Theta^{\bar{a}}{}_{\bar{a}} = \Theta^{\bar{b}}{}_{\bar{b}} \,\,\, \forall \, a,b \,\,\, ; \qquad G^{\bar{c}}{}_{\bar{a}\bar{b}11} = 0 \,\, , \,\, \bar{c} \neq \bar{a}, \bar{b} \,\, .
\ee
Equations (\ref{G-fl-cond}) imply that $\Theta_{lm}$ is completely determined by $\alpha$ and hence there are only two flux parameters: $\alpha$ and $\beta_l$. This is due to the special metric ansatz (\ref{metr}) in which the deformation of the Calabi-Yau three-fold is conformally Calabi-Yau. For a generic deformation, resulting in a non-K\"{a}hler manifold, all three kinds of flux components will be independent.

As is well-known, solving the supersymmetry conditions alone is not enough to ensure that the equations of motion will be satisfied unless the backgrounds of interest are maximally supersymmetric. Therefore, in our case we have to consider also the Bianchi identity and field equation of the four-form $G$ (the Einstein equation is guaranteed to follow from these and supersymmetry \cite{GP}). From (\ref{S_0}) and (\ref{S1}), the $G$ field equation is:
\be
d*G = \frac{1}{2} G \wedge G + 1152 \,(2\pi)^4 \ve X_8 \, ,
\ee
where $\ve$ is the same ${\cal O}(\kappa^{4/3})$ constant as in (\ref{Pa}) and $X_8$ is the eight-form made up of four Riemann tensors that multiplies $C_{(3)}$ in $S_1$ (see (\ref{S1})). We are looking for solutions that preserve the Poincare invariance of the four-dimensional external space and so only components of $G$ along the seven internal directions can be non-vanishing. Therefore the eight-form $G\wedge G = 0$ in the backgrounds of interest. Also, to ${\cal O}(\kappa^{4/3})$ $X_8$ consists of the curvatures for the direct product $M_4\times CY(3)\times \bb{S}^1/\bb{Z}_2$. Writing this space as $\bb{R}^{1,2}\times Y$, where $Y=\bb{R}\times CY(3)\times \bb{S}^1/\bb{Z}_2$, we see that $X_8(\bb{R}^{1,2}\times Y)=X_8(Y)$ due to $\bb{R}^{1,2}$ being flat. Now, recall that up to a numerical coefficient $X_8$ equals $p_2 - \frac{1}{4} p_1^2$, where $p_i$ is the $i$-th Pontryagin class.\footnote{Strictly speaking, topological invariants are mathematically well-defined only for compact manifolds. But since for our purposes we are only interested in their differential form representation, we can disregard that technicality.} In addition, on an eight-manifold admitting a nowhere vanishing spinor the combination $p_2 - \frac{1}{4} p_1^2$ is proportional to the Euler class \cite{IPW}. And since $Y$ is a direct product of a six-manifold and two flat dimensions, the Euler class vanishes identically. Recapitulating, in our case $X_8 = 0$ too.\footnote{This conclusion will most certainly change at higher (than $\kappa^{4/3}$) orders when the $X_8$ term starts feeling the warped background.}

So we are left with the same field equation $D^I G_{IJKL} = 0$ and Bianchi identity, considered in \cite{EW}. Although that work studied the linearized approximation (i.e. ${\cal O}(\kappa^{2/3})$), the part of its analysis that we need is still valid.
More precisely, the $G$ field equation and Bianchi identity imply that the fluxes $\alpha$ and $\beta_a$ are constrained by
\be\label{constraints}
D_{11}\beta_a=\frac i4\partial_a \alpha,~~~~D^m\beta_m=0.
\ee
It is worth emphasizing that the above flux relations 
are exact to all orders in $\kappa$ as long as the eleven-dimensional action contains no higher derivative terms. However, when taking into account the $R^4$ term and its superpartner $C\wedge X_8$,  they will be modified beyond ${\cal O}(\kappa^{4/3})$. Note also that, for the warp factor deformation of the metric 
considered in this section, 
we have $D_{11}\beta=\partial_{11}\beta$ in (\ref{constraints}).

Let us now turn to the analysis of the supersymmetry conditions. As in \cite{CK}, we can extract from (\ref{susyrel}) that $\psi=-b/4$. However, the relation $f(x^m,x^{11})=k(x^m,x^{11})+F(x^{11})$ of \cite{CK} (see eq. (2.18) there) {\it can not} be true in the presence of the $R^4$ correction. Instead, we have
\be \label{kb}
k(x^m,x^{11}) = -2b(x^m,x^{11}) \, ,
\ee
where we have set to zero the undetermined function of $x^{11}$ by a reparametrization of the eleventh coordinate. At this stage one is left with two warp factors $f$ and $b$. As observed in \cite{CK}, the equations for them in (\ref{susyrel}) are generically incompatible. The compatibility conditions arise from $\partial_{a}\partial_{11}f=\partial_{11}\partial_a f = 0$ and a similar equation for $b$:
\be\label{compatibility}
\partial_{11}\bigg(e^{b-f}\beta_a\bigg)=\partial_a\bigg(e^{-b-2f}\alpha\bigg)=0 \, .
\ee
Without the higher derivative correction (i.e. dropping the $Q$ term in (\ref{susyrel})), nontrivial solutions exist {\it only} for $\alpha=0$, $\beta_a\neq 0$ or for $\alpha\neq 0$, $\beta_a = 0$. This can be seen as follows.
From
\bea
\partial_a\alpha&=&
\alpha\partial_a(b+2f)=\frac{i\sqrt 2}{2}e^{b-f}\beta_a\alpha
\nonumber\\
\partial_{11}\beta_a&=&-\beta_a\partial_{11}(b-f)=
\frac{\sqrt{2}}{{12}}e^{-b-2f}\beta_a\alpha\, ,
\eea
and $\partial_{11}\beta_a=\frac{i}4\partial_a\alpha$, we find
\be
i\frac{\sqrt 2}8\beta_a\alpha\bigg
(e^{b-f}+\frac{2}{3}e^{-b-2f}\bigg)=0\,.
\ee
The conclusion is that either $\alpha$ or $\beta_a$ has to vanish in order to have a warp-deformed supersymmetric solution. Notice the essential role in the above argument of the flux constraint (\ref{constraints}), which is a consequence of the field equations (the latter were not considered in \cite{CK}).

Let us now reinstate the correction to the integrability conditions 
(\ref{compatibility}) due to the higher derivative terms (which also means that we have to truncate to order $\kappa^{4/3}$):
\bea
\partial_{11}\beta_a+\frac {\sqrt 2}{12}\alpha\beta_a=0\nonumber\\
\partial_a\alpha+\alpha(\frac{i\sqrt 2}2\beta_a-4\ve \partial_a Q)=0
\label{compexp}\,.
\eea
Notice that in (\ref{compexp}) the term induced by the $R^4$-modified
supersymmetry gravitino variation is, in fact, 
of higher order than the accuracy that we are working at, since 
$\alpha \ve \sim\kappa^{(2+4)/3}$.

We proceed to investigate in turn what happens with the  backgrounds found by \cite{CK}, corresponding to 
either $\alpha=0$ or $\beta_a=0$, in the presence of the $R^4$ terms. 
 
Taking $\alpha=0$, $\beta_a\neq 0$ we see that all 
$x^{11}$-dependence disappears and one can write
\be \label{pbk}
8 \psi(x^m) = -2b(x^m) = k(x^m) \,
\ee
as in \cite{CK}. However, the answer for $f$ is different now. Namely, $\pd_{11} f = 0 = \pd_{11} b$ and $\pd_a f = - 2 \pd_a b - 2 \ve \pd_a Q$ imply that
\be \label{f}
f = - 2b - 2 \ve Q \, .
\ee
Using (\ref{pbk}), (\ref{f}) and equation $-2 \pd_a b = i \frac{\sqrt{2}}{3} e^{k/2-f} \beta_a$ in (\ref{susyrel}), one finds the same relation between warp factor and flux as in \cite{CK} thus recovering the weakly coupled heterotic string result.\footnote{In doing this one has to remember that we are working with accuracy to order $\kappa^{4/3}$ and so should discard higher orders in $-2 \pd_a b = i \frac{\sqrt{2}}{3} e^{k/2-f} \beta_a$, in particular $e^{2\ve Q}\beta_a \approx \beta_a$.} The novelty, due to the $R^4$ correction, comes when one considers the volume of the Calabi-Yau manifold:
\be
V_{CY} = \int e^{3f} \sqrt{g} d^6 x\, .
\ee
Using (\ref{f}), we obtain
\be \label{VCY}
V_{CY} = \int d^6 x \sqrt{g} \,e^{-6b} (1- 6 \ve Q + ...) = V_{CY}^0 - 6 \ve \chi + ... \, ,
\ee
where we have kept only terms up to first order in 
$\ve\sim {\cal O}(\kappa^{4/3})$. 
For the same reason the term with the Calabi-Yau Euler number 
$\chi$ in (\ref{VCY}) is not multiplied by $e^{-6b}$ (recall that $b \sim {\cal O}(\kappa^{2/3})$ and higher). 

Equation (\ref{VCY}) shows explicitly that the $R^4$ correction to the effective action induces a shift of the Calabi-Yau volume as argued in \cite{BG} based on a field redefinition of the dilaton. It was hoped in \cite{CK2} that the same phenomenon might resolved the curvature singularity which appears in their non-linear background at a point along the interval direction at which the warp factor vanishes. We will see below that things are not that straightforward with regard to the proposed singularity resolution mechanism and also that there are other implications. In particular, it will turn out that solutions exist with both $\alpha \neq 0$ and $\beta_a \neq 0$.

Let us now consider the case $\alpha\neq0$, $\beta_a=0$, which is exactly the kind of flux that the strongly coupled non-linear background of \cite{CK} has.
Clearly, we can make the identification
\be
4 \psi (x^{11}) = -b (x^{11})\, ,
\ee
as in \cite{CK}. However, since $\pd_a k = 0$ whereas $\pd_a f \neq 0$ the relation $f = k$ that the background considered in \cite{CK} 
satisfies {\it can not} be true once the higher derivative correction is taken into account. We should note though, that for the present case, i.e. $\beta_a = 0$, equations (\ref{susyrel}) do not imply anything about the relationship between $b(x^{11})$ and $k(x^{11})$. Actually, they leave $k(x^{11})$ completely arbitrary and so by a suitable redefinition of $x^{11}$ one can set $k(x^{11}) = -b(x^{11})$ as in \cite{CK}. On the other hand, we find it most natural to redefine $x^{11}$ such that $e^{k(x^{11})}=1$. Clearly, this freedom has no bearing on the physics of the solution. Specializing the relevant equations in (\ref{susyrel}) to the present case, we find
\be \label{fsol}
f(x^m,x^{11}) = -2\ve Q(x^m)-b(x^{11}) \, .
\ee
Hence expanding to order $\kappa^{4/3}$ the equation that determines the $x^{11}$ dependence of $f$,
\be \label{feq}
\pd_{11} f = -\frac{\sqrt{2}}{24} e^{k/2-2f} \alpha \, ,
\ee
we end up with
\be
e^{2f(x^m,x^{11})} =1 -\frac{\sqrt{2}}{12}\int_0^{x^{11}} dz  \,\, \alpha (z) -4\ve Q(x^m) +{\cal O}(\kappa^{6/3}) \, .
\ee
The above solution is consistent since when $\beta_a=0$ the constraints (\ref{constraints}) imply $\partial_a \alpha=0$.
Taking the flux sources to be localized at the two ends of the interval, the Bianchi identity is solved by $G_{(4)}=\Theta(x^{11}) {\cal S}_{(4)}(x^m)$, where ${\cal S}_{(4)}(x^m)$ is a closed four-form. The corresponding value of 
the $\alpha$ flux is: 
\be
\alpha=\Theta(x^{11}){\cal S}_{mnpq}J^{mn}J^{pq}=4\Theta(x^{11}){\cal S}_{a\bar{b}c\bar{d}}J^{a\bar{b}}J^{c\bar{d}}\equiv 4\sqrt{2}\,\Theta(x^{11}){\cal S}\, .
\ee
So we find for the volume of the Calabi-Yau:
\be\label{vcy}
V_{CY}(x^{11})=\int d^6 x \sqrt{g} e^{3f} = V_{CY}^0\bigg
(1-x^{11}\Theta(x^{11}){\cal S}+
\frac{1}{6}(x^{11}\Theta(x^{11}){\cal S})^2\bigg)-6\ve \chi \, .
\ee
We see again the expected constant shift proportional to the Euler number.

Finally, let us note that we can construct to the same level of accuracy 
solutions which have both $\alpha\neq 0$ and $\beta_a\neq 0$,  for example by taking $\beta=\beta(x^m)\sim\kappa^{4/3}$ meaning also that $\alpha= \alpha(x^{11})$. In particular, for
\be
\beta_a = i C \ve \partial_a Q \, ,
\ee
where $C$ is an arbitrary real number, one can solve 
both the supersymmetry conditions (\ref{susyrel}) and the flux constraints (\ref{constraints}). Indeed, writing $f(x^a,x^{11})=f_1(x^a)+f_2(x^{11})$ and $b(x^a,x^{11})=b_1(x^a)+b_2(x^{11})$, one finds $b_1 = - 2 f_1$, $b_2 = -f_2$ and: 
\be 
f_1(x^a) = \left(-\frac{\sqrt{2} C}{3} - 2\right) \ve Q \, , \qquad e^{f_2(x^{11})} = e^{f_2(0)} -\frac{\sqrt{2}}{24} \int_0^{x^{11}} dz \, \alpha (z) \, ,
\ee
where we have used the relation $k(x^a, x^{11})=-2b(x^a, x^{11})$, (\ref{kb}), that has to be satisfied when both $\alpha \neq 0$ and $\beta_a \neq 0$.

\subsection{General case}

Let us turn now to the metric ansatz
\bea
&&\!\!\!\!\!\!\!\!\!\!
ds^2 = e^{b(x^m,x^{11})} \eta_{\mu \nu} dx^{\mu} dx^{\nu} + [g_{ln} (x^m) + h_{ln} (x^m,x^{11})] dx^l dx^n + e^{k(x^m,x^{11})}(dx^{11})^2 \nn\\
&&=\hat{g}_{MN} dx^M dx^N \, ,
\eea
which allows more general flux. Since generically the deformation $h_{mn}(x^m , x^{11})$ does not preserve the K\"{a}hler property of the Calabi-Yau metric $g_{mn} (x^m)$, we can not describe the $G$-flux in terms of the same quantities as in (\ref{def}). A convenient new parametrization is given by
\be
G = \hat{g}^{a\bar{b}} \hat{g}^{c\bar{d}} G_{a\bar{b} c\bar{d}} \, , \qquad G_m = \hat{g}^{b\bar{c}} G_{mb\bar{c}11} \, , \qquad G_{mn} = \hat{g}^{c\bar{d}} G_{mnc\bar{d}} \, .
\ee

Splitting the vielbein of the internal six-dimensional space as $\hat{e}^{\underline{l}}{}_n (x^m,x^{11})$ $= e^{\ul{l}}{}_n (x^m) + f^{\ul{l}}{}_n (x^m,x^{11})$, where $e^{\ul{l}}{}_n (x^m)$ is the vielbein for $g_{ln}$, one finds for the spin-connection \cite{CK}:
\bea
&&\Omega_{l \ul{m} \ul{n}} (\hat{e}) = \Omega_{l \ul{m} \ul{n}} (e) + \Omega_{l \ul{m} \ul{n}}^{(d)} (e,f) \, , \qquad \Omega_{11 \ul{l} \ul{11}} (\hat{e}) = -\frac{1}{2} \hat{e}_{\ul{l}}{}^l \hat{e}_{\ul{11},11} \pd_l k \, , \nn\\
&&\Omega_{l \ul{m} \ul{11}} (\hat{e}) = \frac{1}{2} \hat{e}_{\ul{11}}{}^{11} (\pd_{11} f_{\ul{m} l}+\hat{e}_{\ul{m}}{}^m \hat{e}^{\ul{l}}{}_l \pd_{11} f_{\ul{l} m}) \, , \qquad \Omega_{11 \ul{l} \ul{m}} (\hat{e}) = \hat{e}_{[\ul{l}}{}^l \pd_{|11|} f_{\ul{m} ] l} \, , \nn\\
&&\Omega_{\mu \ul{\nu} \ul{l}} (\hat{e}) = \frac{1}{2} \hat{e}_{\ul{l}}{}^m \hat{e}_{\ul{\nu}\mu} \pd_m b \, , \qquad \Omega_{\mu \ul{\nu} 11} (\hat{e}) = \frac{1}{2} \hat{e}_{\ul{\nu} \mu} \hat{e}_{\ul{11}}{}^{11} \pd_{11} b \, .
\eea
In the above formulae $\Omega_{l \ul{m} \ul{n}} (e)$ is the spin-connection of the initial $CY(3)$ whereas $f^{\ul{l}}{}_m$ and $\Omega_{l \ul{m} \ul{n}}^{(d)} (e)$ measure the deviation from it.

It is easy to compute the contribution of the $R^4$ induced term in the gravitino variation to the supersymmetry conditions (7.93)-(7.103) of \cite{CK}. We write down only the single equation that changes:
\be \label{Omd}
\Omega_{ab}^{(d)}{}^b - \Omega_{a \bar{b}}^{(d)}{}^{\bar{b}} = \frac{2\sqrt{2}}{3} e^{-k/2} G_a + 2 \ve \pd_a Q \, .
\ee
As explained in \cite{CK}, one can find the $x^{11}$ dependence of the Calabi-Yau volume by using $\pd_{M} \sqrt{\hat{g}} = \sqrt{\hat{g}} \,\Gamma^N{}_{NM} (\hat{g})$. Relating $\Gamma^N{}_{N11}$ to $\Omega^N{}_{N11}$ via the vielbein postulate one finds:
\be \label{V-CY}
\pd_{11} \sqrt{\hat{g}_{CY}} = \sqrt{\hat{g}_{CY}} \,\Omega^m{}_{m11} (\hat{e}) \, .
\ee
Since the higher derivative correction appears only in the equation for $\Omega^{(d)}_{lmn}$ but not those for $\Omega^m{}_{n11}$ (or any of the warp factors $b,k,\psi$), we conclude from (\ref{V-CY}) that this correction does not  affect the $x^{11}$ dependence of $V_{CY}$. This is consistent with the constant shift of the Calabi-Yau volume that it was inducing for the special flux of the previous subsection. Unfortunately though, in the present general case one can not be more explicit than this.\footnote{Similarly to $\pd_{11} \sqrt{\hat{g}_{CY}}$, one can consider $\pd_{a} \sqrt{\hat{g}_{CY}}$ but the expression for it contains $\Omega_{ab}^{(d)}{}^b + \Omega_{a \bar{b}}^{(d)}{}^{\bar{b}}$ instead of $\Omega_{ab}^{(d)}{}^b - \Omega_{a \bar{b}}^{(d)}{}^{\bar{b}}$ as in (\ref{Omd}) and so no definite statement can be made about the role of the $\ve \pd_a Q$ term.} 

The search for deformed backgrounds allowing generic flux may be significantly facilitated by the $G$-structures approach. So far we have not been able to find new solutions, but in Appendix B we present some preliminary considerations based on the results of \cite{DP}. More precisely, we derive generalized Hitchin flow equations for non-K\"{a}hler six-manifolds fibered over an interval.\footnote{These considerations assume that the internal manifold of the heterotic string compactification has $SU(3)$ structure. However, in the presence of higher derivative corrections to the effective action the internal manifold might even be noncomplex (see the discussion in Section 4 of \cite{Lust1}). This case would be very difficult to analyze and we have nothing to say about it at present.}

As a last remark in the current section, we note that the change in $\Omega^{(d)}$, implied by equation (\ref{Omd}), means that the $R^4$ correction has an impact on the holonomy group ${\cal H}$ of the deformed (generically non-K\"{a}hler) six-dimensional manifold since ${\cal H}$ is generated by parallel transport, determined by the full spin-connection $\Omega (\hat{e}) = \Omega (e) + \Omega^{(d)} (e,f)$, around all closed curves in the manifold.

\section{Scalar potential}
\setcounter{equation}{0}

Let us now combine ingredients from previous sections to address the role of the $R^4$ correction in shaping the scalar potential of the four-dimensional effective theory. More precisely, we will be interested in the fate of the de Sitter vacua of \cite{BCK}.

It may seem that such questions are out of reach if one uses only a finite number of terms in a perturbative expansion, as we have done. However, recall that the minima of \cite{BCK} occur for $\epsilon = \frac{2 \pi L}{3 V_v^{2/3}} (\frac{\kappa}{4 \pi})^{\frac{2}{3}} \sim {\cal O}(1)$ and even $\epsilon < 1$, which is just at the border of validity of the perturbative expansion. In addition, since higher derivative terms in M-theory like $R^4$, the higher order $D^4R^4$ \cite{GKV}\footnote{The notation $D^4R^4$ is symbolic; the index contractions are different from those in $R^4$ or $X_8$.} etc. change the supersymmetry conditions and field equations (in particular, the constraints (\ref{constraints})), it is not a priori clear whether the better approximation is to extend to all orders a solution that neglects them, as in \cite{CK}, or to keep only terms up to the appropriate order of accuracy (for us ${\cal O}(\kappa^{4/3})$). The latter option merits an investigation on its own, which we now turn to.

The scalar potential in $N=1$ $d=4$ supergravity is given by
\be \label{ScPot}
U = e^{K} \left( K^{A \bar{B}} D_A W D_{\bar{B}} \overline{W} - 3 |W|^2 \right) + U_D\, ,
\ee
where $K$ is the K\"{a}hler potential, $W$ - the superpotential, $K^{A \bar{B}}$ - the inverse moduli space metric i.e. the inverse of $K_{A \bar{B}} \equiv \pd_A \pd_{\bar{B}} K$, the derivative $D_A W \equiv \pd_A W + K_A W$ and $U_D$ are D-terms for the charged matter fields $C^p$ that originate from the gauge multiplets localized on the boundaries of the eleven-dimensional space-time. In principle, the indices $A,B$ in (\ref{ScPot}) range over all scalar fields in the effective action. However, as in \cite{BCK} we do not address here the stabilization of the Calabi-Yau complex structure moduli and the vector bundle moduli.\footnote{The latter have been shown to be significantly suppressed \cite{BO}.} Presumably this can be achieved similarly to the weakly coupled case and the novelty is in the stabilization of the orbifold length. Again as in \cite{BCK}, we take $h^{1,1} = 1$. One can show that in the present case too the matter fields $C^p$ can be stabilized at a nonzero but strongly suppressed value and so they can be safely neglected for the purposes of minimization of the scalar potential $U$ w.r.t. to the remaining fields, which we do from now on.

The $R^4$ corrections in the effective action change the K\"{a}hler potential, but not the superpotential.\footnote{Indeed, in our derivation of the K\"{a}hler potential in Appendix A we have only neglected terms with more than two derivatives but not terms that would contribute to the superpotential. That such do not arise from the $R^4$ term is clear from the fact that each curvature component in (\ref{curv}) contains two derivatives.}  Recall that one expects higher derivative terms to not modify the superpotential for reasons similar to the original argument of \cite{EW2} about $\alpha'$ corrections in string theory. Namely, the Peccei-Quinn symmetries of the low energy theory (the shift symmetries of the axions $\sigma_S$ and $\sigma_T$ in (\ref{s_ph})), that are inherited from the three-form gauge transformation $C \rightarrow C + d\omega$ with $\omega$ being a 2-form, are broken only by non-perturbative effects, like M2-brane instantons wrapping topologically non-trivial 3-cycles \cite{HM}. Whereas, if higher derivative terms contributed $S,T$ dependence to the superpotential, that would break perturbatively the axionic shift symmetries.

From (\ref{del_K}) we see that the change in the K\"{a}hler potential, due to the $R^4$ corrections, induces the following change in the scalar potential to ${\cal O}(\kappa^{4/3})$:
\be
\delta U =b_2\kappa^{4/3}\chi\bigg( -\frac{1}{6 (S + \bar{S})}U_0 
+\frac{1}6e^{K_0}(\overline W D_S^0 W+W \overline{D_S^0 W})+
\frac{(S+\bar S)}3 e^{K_0}
D_S^0 W \overline{D_S^0 W}\bigg) \, ,
\label{deltau}
\ee
where 
$K_0 = - \ln (S+\bar{S}) - 3 \ln (T + \bar{T})$, $D^0_S = \pd_S + K_{0,S}$ and 
$U_0 (K,W) = U(K_0,W)$. 
Note that, although the K\"{a}hler potential for the superfield $T$ does not change, the potential for it does, as the superpotential is a function of both $S$ and $T$: $W=W(S,T)$. 
The non-perturbative part of the superpotential $W_{\rm{non-pert}}$ 
is given by the sum of two contributions: from open membrane instantons, $W_{OM}$, and from gaugino condensation, $W_{GC}$. 

Without the charged matter sector, the perturbative contribution to the 
superpotential vanishes and one is left with \cite{LOPR,LOW2}:  
\be
W=W_{OM}+W_{GC}=he^{-T}+g e^{-(S-\gamma T)/C_H}\, ,
\ee 
where $g=-C_H(2M_{GUT}l_{11})^3$ with $C_H$ being the dual Coxeter number of the hidden gauge group. The open membrane instanton Pfaffian $h$ is bounded by $|h|\le (2M_{GUT}l_{11})^3$. Finally, the slope $\gamma$ is determined in terms of the 
the length of the interval $\cal L$ and the flux on the visible boundary ${\cal S}$ via $\gamma={\cal  L S}$. For convenience and easier comparison with \cite{BCK}, from now on we introduce the notation:
\be \label{new_def_ST}
S={\cal V}+i\sigma_S \, ,~~~~~T={\cal V}_{OM}+i\sigma_T \, ,
\ee
where ${\cal V}={\cal V}({\cal L})$ is the average volume of the Calabi-Yau and 
${\cal V}_{OM}={\cal V}_{OM}({\cal L})$ is the volume of the open membrane instanton:
\be\label{avvol}
{\cal V}({\cal L})=\frac{1}{{\cal L}}\int_0^{L} dx^{11}e^{k/2}
 V_{CY}(x^{11})~~~~,~~~~
{\cal V}_{OM}({\cal L})={\cal L}\cV^{1/3}({\cal L})\, .
\ee
We differ from \cite{BCK} in our averaging over the orbifold direction; 
by introducing the measure $dx^{11} e^{k/2}$, and defining the orbifold length with respect to the same measure 
${\cal L}=\int_0^{L}dx^{11}e^{k/2}$, we ensure that the average volume 
is independent of redefinitions of the $x^{11}$ coordinate (or equivalently, $\cV ({\cal L})$ is the same irrespective of the freedom we noted in Subsection 4.1 in choosing the value of the warp factor $e^{k(x^{11})}$). We should also 
recall that, whereas \cite{BCK} employed a ``fully'' non-linear
background derived by ignoring all higher derivative corrections to the eleven-dimensional supergravity action, we must perform an expansion in $\kappa$. It follows from the previously derived Calabi-Yau volume dependence on $x^{11}$ (\ref{vcy}) that, to the order of accuracy we are working to, the average volume of the Calabi-Yau is:
\be
\cV=\cV_v\left[\bigg(1-\frac{1}{2}{\cal L S}+\frac{
1}{18}{\cal L}^2{\cal S}^2\,\bigg) -6\ve\chi\right]\, .
\ee    
where we denoted, as before, $\varepsilon=b_1 T_2\kappa^2$.

For the zeroth order (direct product $CY(3)\times \bb{S}^1/\bb{Z}_2$) metric, (\ref{new_def_ST}) reduces to $S=e^{6a}+i\sigma_S$ and $T=e^{c+2a}+i\sigma_T$. But the warping of the strongly coupled background introduces ${\cal L}$-dependence of the Calabi-Yau volume ${\cV}$. As in \cite{BCK}, we take for the modulus $\cV_v$ (the volume of the Calabi-Yau at the visible boundary) a value dictated by phenomenology and study the resulting minima for ${\cal L}$.\footnote{The stabilization of ${\cV_v}$ could presumably be achieved by compactification on a 6d non-K\"{a}hler manifold \cite{nonK,BCK}.} 
The existence of a minimum for the orbifold length
${\cal L}$ is guaranteed, as observed in \cite{BCK}, by the two 
opposing nonperturbative effects: the runaway behavior
of the scalar potential, due to open membrane instantons, is being offset by the contribution coming from gaugino condensation on the hidden boundary.

The scalar potential, including the $R^4$ induced correction (\ref{deltau}),
is given by
\bea \label{U}
U&=&\frac{1}{2^4{\cal V}_{OM}^3{\cal V}}\bigg(
\frac{1}{C_H^2}|W_{GC}|^2
(4{\cal V}^2+\frac{b_2\chi{\cal V}}3)+\bigg|\frac\gamma{C_H}W_{GC}-W_{OM}\bigg|^2(\frac{4{\cal V}^2_{OM}}3-
\frac{b_2\chi{\cal V}_{OM}^2}{9{\cal V}})\nonumber\\
&+&|W_{OM}+W_{GC}|^2
(1-\frac{b_2\chi}{12{\cal V}})
-\frac{1}{C_H}(W_{GC}
(\overline{W_{GC}}+\overline{W_{OM}})+c.c.)(-2{\cal V})
\nonumber\\
&+&((\frac{\gamma}{C_H}W_{GC}-W_{OM})
\overline{(W_{OM}+W_{GC})}+c.c.)(-2{\cal V}_{OM}+\frac{b_2\chi{\cal V}_{OM}}{6{\cal V}})
\bigg)\, .
\eea
All volumes and lengths are defined to be dimensionless 
by appropriately factorizing the eleven-dimensional Planck scale, $l_{11}$. For the same reason, 
we dropped the $\kappa^{4/3}$ factor from $\delta U$.  
The scalar potential (\ref{U}) can be written explicitly in terms of 
the moduli as
\bea
U\!\!\!\!&=&\!\!\!\!\frac{1}{2^4}\bigg[
g^2e^{-2({\cal V}-\gamma{\cal V}_{OM})/C_H}\bigg(\!\frac{4{\cal V}^2}{C_H^2}+
\frac{4\gamma^2{\cal V}_{OM}^2}{3C_H^2}+\frac{4{\cal V}}{C_H}-\frac{4\gamma{\cal V}_{OM}}{C_H}
+\frac{b_2\chi{\cal V}}3-\frac{b_2\chi\gamma^2{\cal V}_{OM}^2}{9C_H^2{\cal V}}\!\bigg)
\nonumber\\
&+&\!\!\!\!|h|^2e^{-2{\cal V}_{OM}}\bigg(\frac{4{\cal V}_{OM}^2}3+4{\cal V}_{OM}-
\frac{b_2\chi{\cal V}_{OM}^2}{9{\cal V}}\bigg)\nonumber\\
&-&\!\!\!\!2ge^{-{\cal V}_{OM}-({\cal V}-\gamma{\cal V}_{OM})/C_H}
\left(\!Re(h)\cos(\frac{\sigma_S}{C_H}-(\frac\gamma{C_H}+1)\sigma_T) \right. \nn \\
&+& \!\!\!\!\left. Im(h)\sin(\frac{\sigma_S}{C_H}-(\frac\gamma{C_H}+1)\sigma_T)\right) \!\!\bigg(\!\!\frac{4\gamma{\cal V}_{OM}^2}{3C_H}-\frac{2{\cal V}}{C_H}+\frac{2\gamma{\cal V}_{OM}}{C_H}-
\frac{b_2\chi\gamma{\cal V}_{OM}^2}{9C_H{\cal V}}\bigg)\bigg]\!
\frac{1}{{\cal V}_{OM}^3{\cal V}}\,,\nonumber\\
\label{uuu}
\eea
where we have kept only the leading and next-to-leading 
order terms in an expansion in powers of 
$1/{\cal V}$, $1/{\cal V}_{OM}$.\footnote{It is perhaps
worth noting that (\ref{U}) contains all contributions in $1/\cV$, $1/\cV_{OM}$ of the original potential, i.e. with $\chi=0$.} This truncation is justified since 
near the minimum one must have $\cV >\!\!> 1$ and $\cV_{OM} >\!\!>1$ for the validity of the supergravity approximation to M-theory. Due to the same condition (namely, $\cV_{OM} >\!\!>1$), the contributions of multiply wrapped membranes, which are not well-understood, can be neglected. 

From (\ref{uuu}) one can see that the $R^4$ corrections appear at the next-to-leading order in $1/{\cal V}$, $1/{\cal V}_{OM}$ compared to the leading terms (the first line in (\ref{U}) with $b_2\chi = 0$) that were kept in \cite{BCK}. Hence, non-surprisingly their effect is only a few percentage as we will see below.\footnote{One might have thought that, similarly to the string case \cite{BB}, the $R^4$ corrections would be dominant if the complex structure moduli were assumed to be stabilized at a supersymmetric minimum by a classical superpotential, for example from compactifying on a 6d non-K\"{a}hler manifold. This does not happen however, due to the fact that the no-scale structure of the K\"{a}hler potential for the $T$-modulus is preserved in our case.} The main reason for the differences we find is in the expansion of the background to order $\kappa^{4/3}$.

Extremizing with respect to the axionic scalars $\sigma_{S,T}$ 
yields the same condition as in \cite{BCK}
\be\label{axsc}
{\rm tan}(\frac{\sigma_S}{C_H}-(\frac\gamma{C_H}+1)\sigma_T)=
\frac{Im\,h}{Re\,h}\, .
\ee
Substituting the latter in (\ref{uuu}), to leading order in the expansion in powers of $1/{\cal V}$ and $1/{\cal V}_{OM}$, the scalar potential is a sum of squares:
\be
U\!\!=\!\!\frac{1}{2^4\cV\cV_{OM}^3}\bigg[\frac{4\cV_{OM}^2}3
\bigg(\!\!\frac{\gamma}{C_H}g
e^{-(\cV-\gamma\cV_{OM})/C_H}\!-(\pm)
|h|e^{-\cV_{OM}}\!\!\bigg)^{\!\!2}\!\!+\frac{4{\cal V}^2}{C_H^2}
g^2e^{-2({\cal V}-\gamma{\cal V}_{OM})/C_H}\!\bigg]\!\!+\!... \,\, .\label{sumsq}
\ee 
This positivity of the vacuum energy led the authors of \cite{BCK} to the conclusion that meta-stable\footnote{Clearly, the global minimum is given by $U=0$ and is obtained in the decompactification limit $\cV \rightarrow \infty$, $\cV_{OM} \rightarrow \infty$.} de Sitter vacua exist in heterotic M-theory. The two sign choices in (\ref{sumsq})
follow from solving for the axionic scalars from the extremum condition
(\ref{axsc}). 

Keeping the sub-leading terms of the expansion in $1/{\cal V}$, $1/{\cal V}_{OM}$ as well as the
$R^4$ induced corrections, we proceed next to extremize the scalar potential
with respect to the orbifold length.
The equation $\pd U/ \pd {\cal L} = 0$ can not be solved analytically. 
We have studied its solutions numerically for the
same values of the various parameters as in \cite{BCK}: $|h|=10^{-8},
\cV_v=800$, $C_H=8$ corresponding to a hidden gauge group $SO(10)$, 
${\cal S}=(\frac{6}{10\cV_v})^{1/3}$. 
 The $R^4$ induced correction is proportional to the Euler number of the Calabi-Yau manifold. 
\begin{figure}
\centerline{\psfig{figure=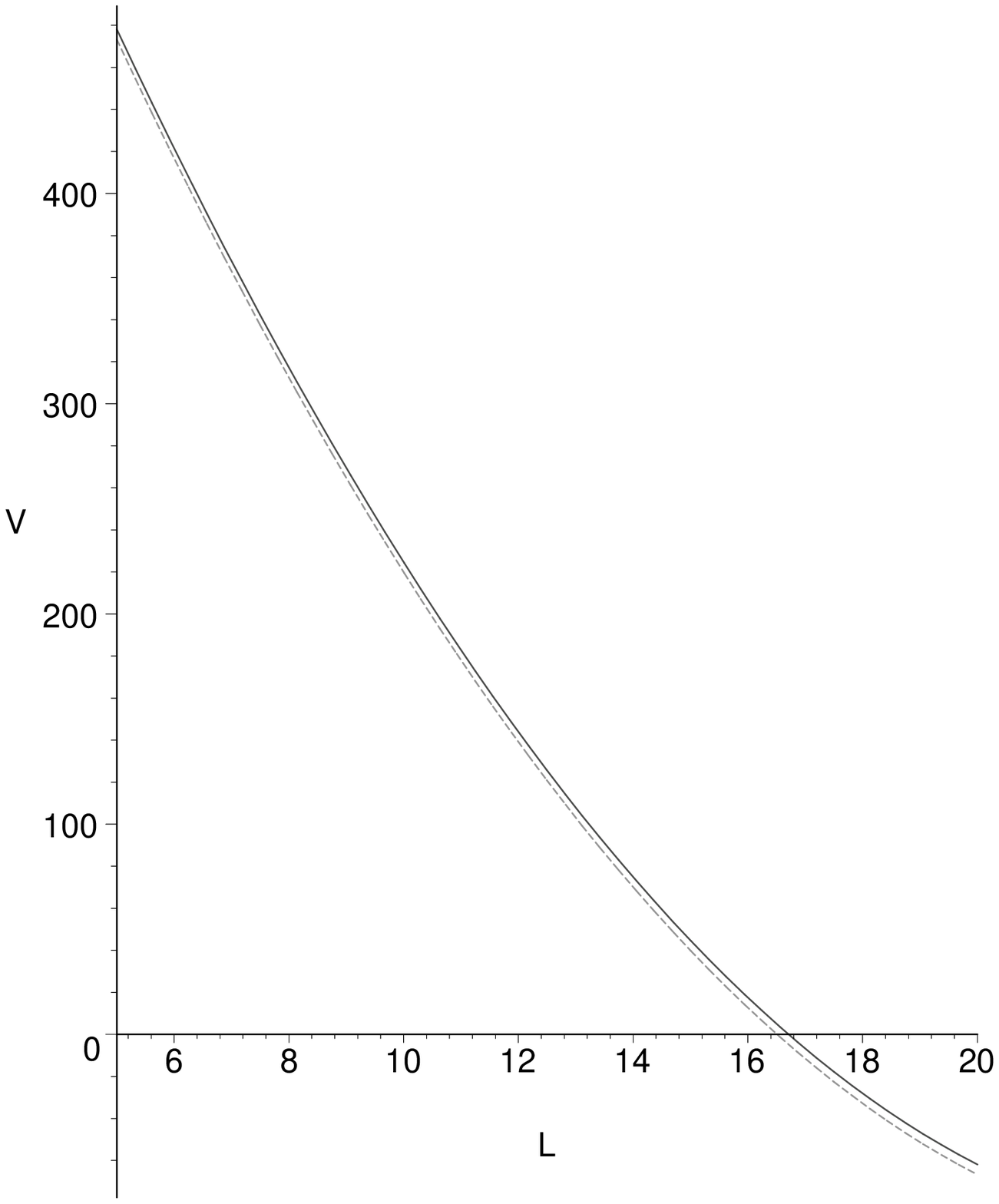,width=6cm,clip=}
~~~~~~~~~~~~~~~~\psfig{figure=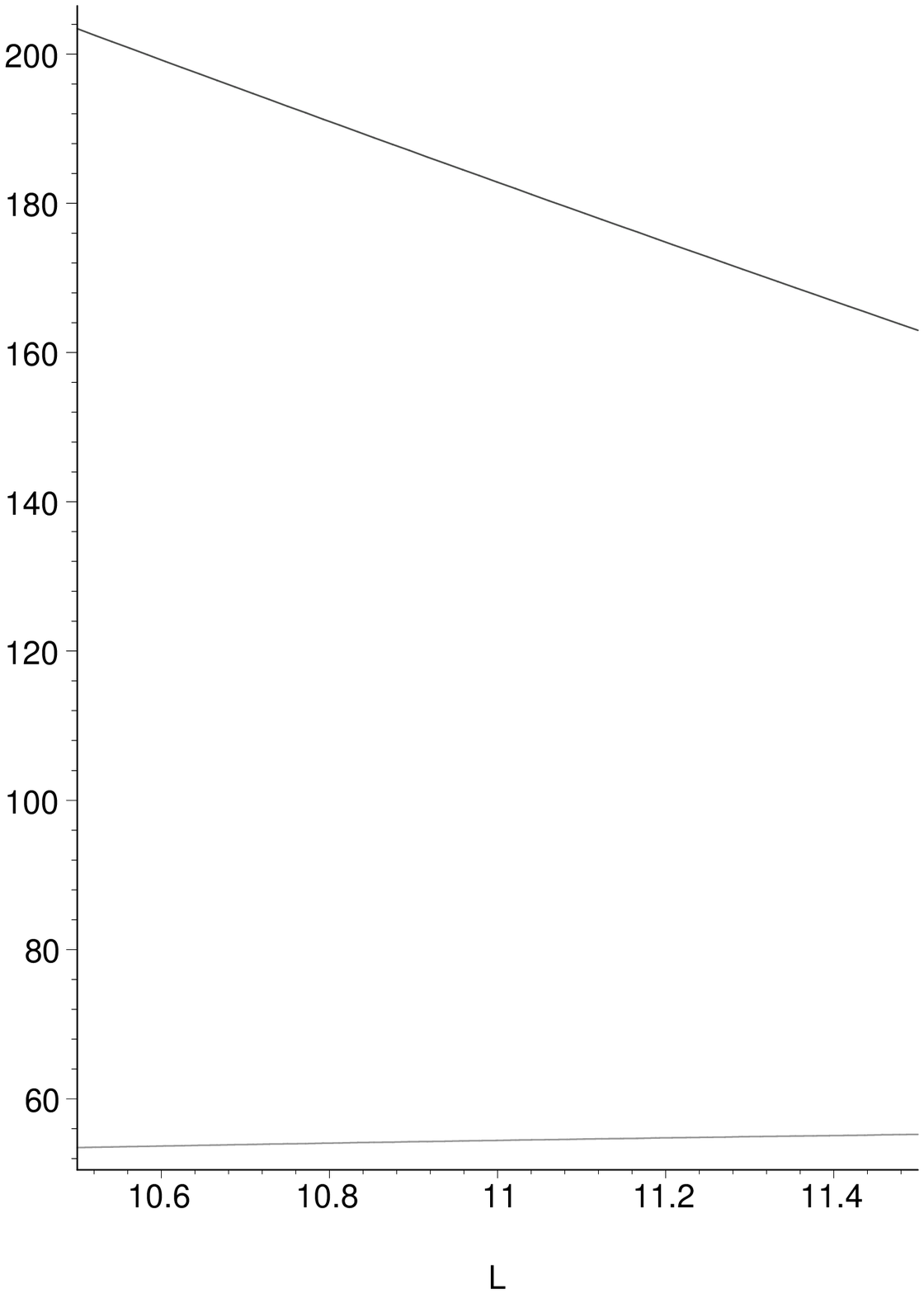,width=5cm,clip=}}  
\caption{\small 
The average volume of the Calabi-Yau $\cV$ as a function of the orbifold 
length ${\cal L}$: 
the solid line represents the $R^4$-corrected volume for $\chi<0$,  
while the dotted line corresponds to $\chi=0$. Also on the right are 
depicted the volumes $\cV, \,\cV_{OM}$ as a function of the orbifold length 
in the vicinity of the extremum.  }
\label{} 
\end{figure}
\noindent 

Figure 1 depicts the average Calabi-Yau volume $\cV$ as a function of the orbifold length ${\cal L}$ for the $\kappa^{4/3}$ expansion of the ``exact'' background of \cite{CK, BCK}. The solid line corresponds to a negative Euler number, whereas the dotted one - to $\chi = 0$ meaning also no $R^4$ correction. 
As it is transparent from the second plot of Figure 1, both volumes
$\cV,\,\cV_{OM}$ are sufficiently large near the minimum to justify the use of 
supergravity as an effective action.
At the same time, the condition $\cV >\!\!> \cV_{OM}$ is satisfied to a 
reasonable degree, which justifies the use of the perturbative expansion 
in $\kappa^{2/3}$. 

We would also like to comment on the positivity of the volume of the 
Calabi-Yau: from the analysis of \cite{CK}, it might be construed 
that the volume is positive for all values of $x^{11}$ 
as a consequence of using the ``fully'' non-linear background. With
the choice of the warp factor $k(x^{11})=f(x^{11})$, possible in the 
absence of the 
higher derivative corrections as discussed extensively in Section 4.1,
the authors of \cite{CK} have found that the volume of the Calabi-Yau
is equal to $V_{CY}^0=\sqrt{e^{6f}}=
\bigg[(1-x^{11}\Theta(x^{11}){\cal S})^{2/3}\bigg]^3$. Hence,
$V_{CY}^0=(1-x^{11}\Theta(x^{11}){\cal S})^2$, and using this expression of the volume, the authors of \cite{CK} solved for the warp factors in terms of roots of 
a quantity that is manifestly positive, $V_{CY}^0$. 
We believe this to be misleading since the warp factors were derived first,
using the Killing spinor equations: $e^{3/2 f}=1-x^{11}\Theta(x^{11}){\cal S}$ 
according to eq (5.47) in \cite{CK}.
This implies that $x^{11}$ cannot be defined in this coordinate system beyond 
$x^{11}_{\rm{max}}=1/{\cal S}$, otherwise the warp factors will become 
negative.
Moreover, the positivity of the volume of the CY for all values of 
$x^{11}$ is a coordinate dependent statement: with $k=0$, for instance, 
one finds $V_{CY}^0=(1-x^{11}\Theta(x^{11}){\cal S})^{3/2}$, which carries
the same implications, namely $x^{11}\leq 1/{\cal S}$.  On the other hand,
the average volume of the Calabi-Yau (\ref{avvol}), which depends only on the 
orbifold length ${\cal L}$, is coordinate
independent, i.e. it is the same function $\cV({\cal L})$ 
irrespective of the choice of the warp factor $k(x^{11})$.

In Figure 2, we plotted the scalar potential near its minimum as 
a function of the orbifold length, for both signs arising from the 
extremization with respect to the axionic scalars. For each choice of sign
we plotted both the truncated ${\cal O}(\kappa^{4/3})$ background
scalar potential, and the $R^4$ corrected potential, assuming a negative
Euler number. From Figure 2 we see that still there is a de Sitter minimum, 
although the value of the 
scalar potential at the minimum is much smaller than the one reported 
in \cite{BCK}, which was of order $10^{-58}$ for the same set of parameters. 
Leaving aside the difference in defining the average volume, 
this is due solely to the effect of truncating the exact background to order
$\kappa^{4/3}$, i.e. performing an expansion in the background flux up to 
second order and ignoring terms of order ${\cal O}({\cal S}^3)$ and higher.
Another consequence of this truncation is that the
minimum is shifted towards bigger values of the orbifold length ${\cal L}$.
\begin{figure}
\centerline{\psfig{figure=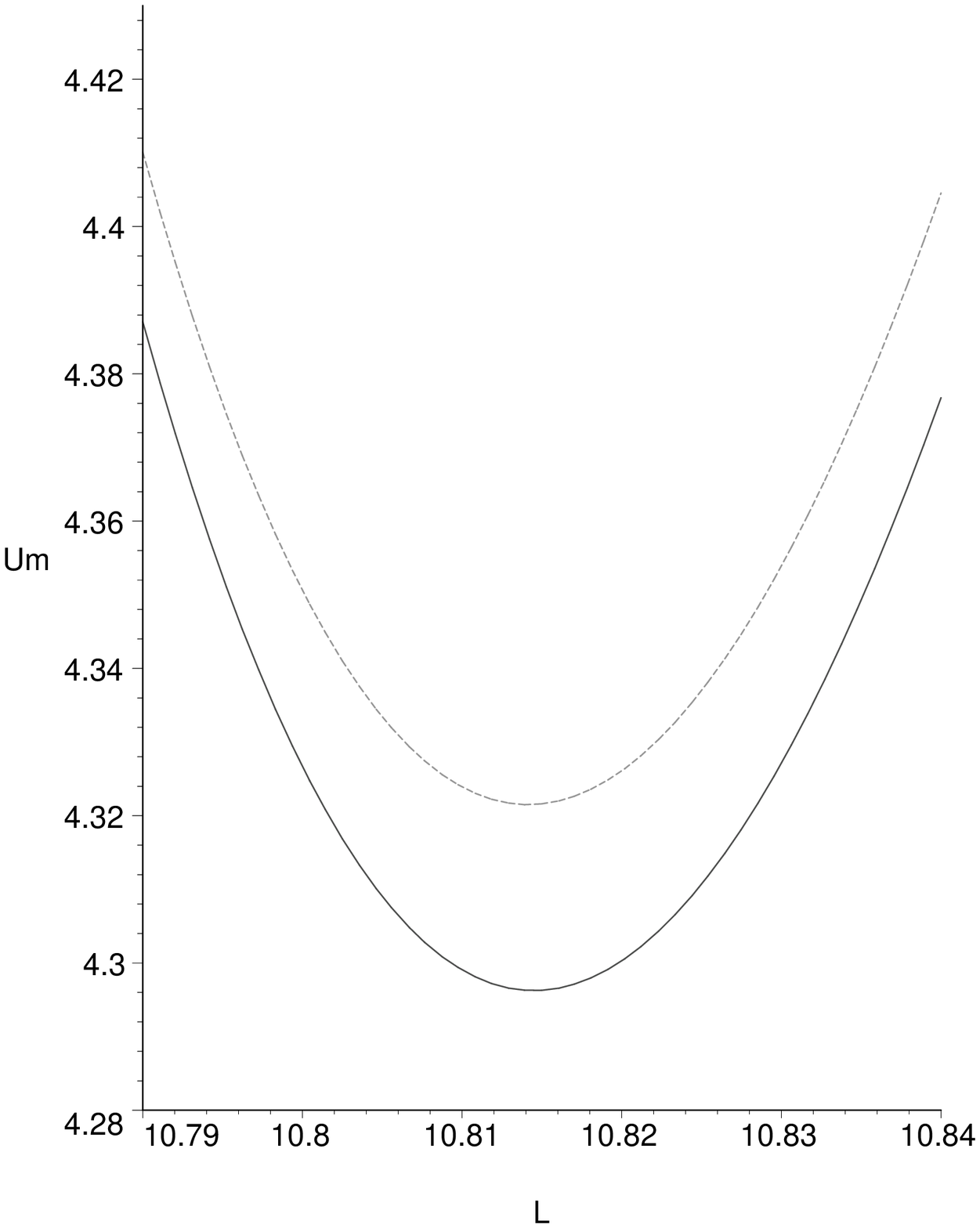,width=6.5cm,clip=}~~~~~~~~~~~~
\psfig{figure=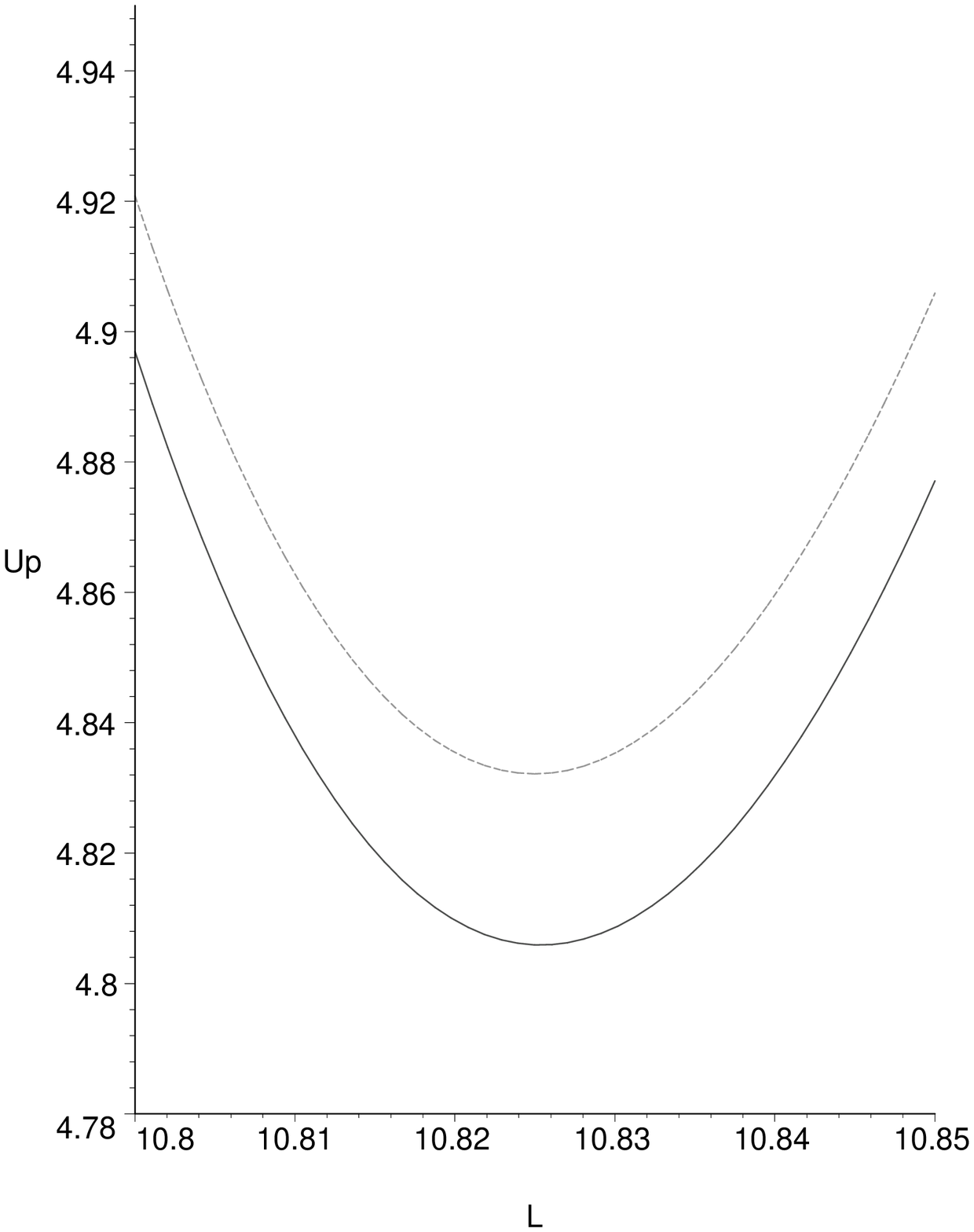,width=6.5cm,clip=}}
\caption{\small{The scalar potential, after the axionic scalar extremization with 
the two possible signs - (Um$\times10^{-91}$) and + (Up$\times10^{-91}$), 
for $b_2\chi=-100$, as a function of the orbifold length ${\cal L}$:
the solid line represents the $R^4$ corrected potential and the dotted line
corresponds to $\chi=0$.}}  
\end{figure} 
Finally, Figure 3 shows the dependence of the scalar potential on the 
Euler number. Clearly, the correction to the volume and that to the 
scalar potential have opposite signs: for negative Euler numbers,
the average volume is increasing, whereas the value of the 
scalar potential at the minimum shifts towards zero.
\begin{figure}
 \centerline{\psfig{figure=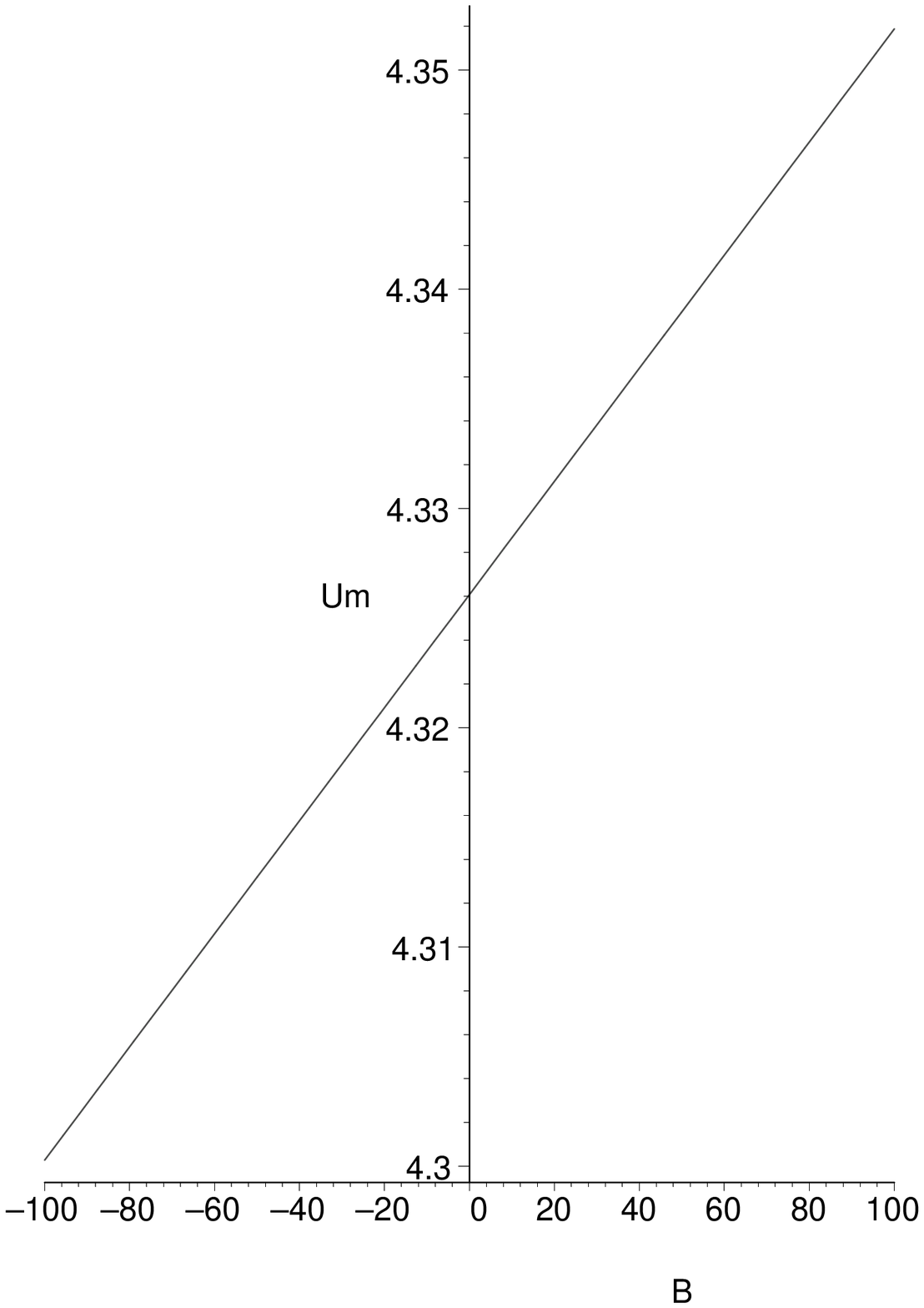,width=5.8cm,clip=} 
~~~~~~~~~~~~\psfig{figure=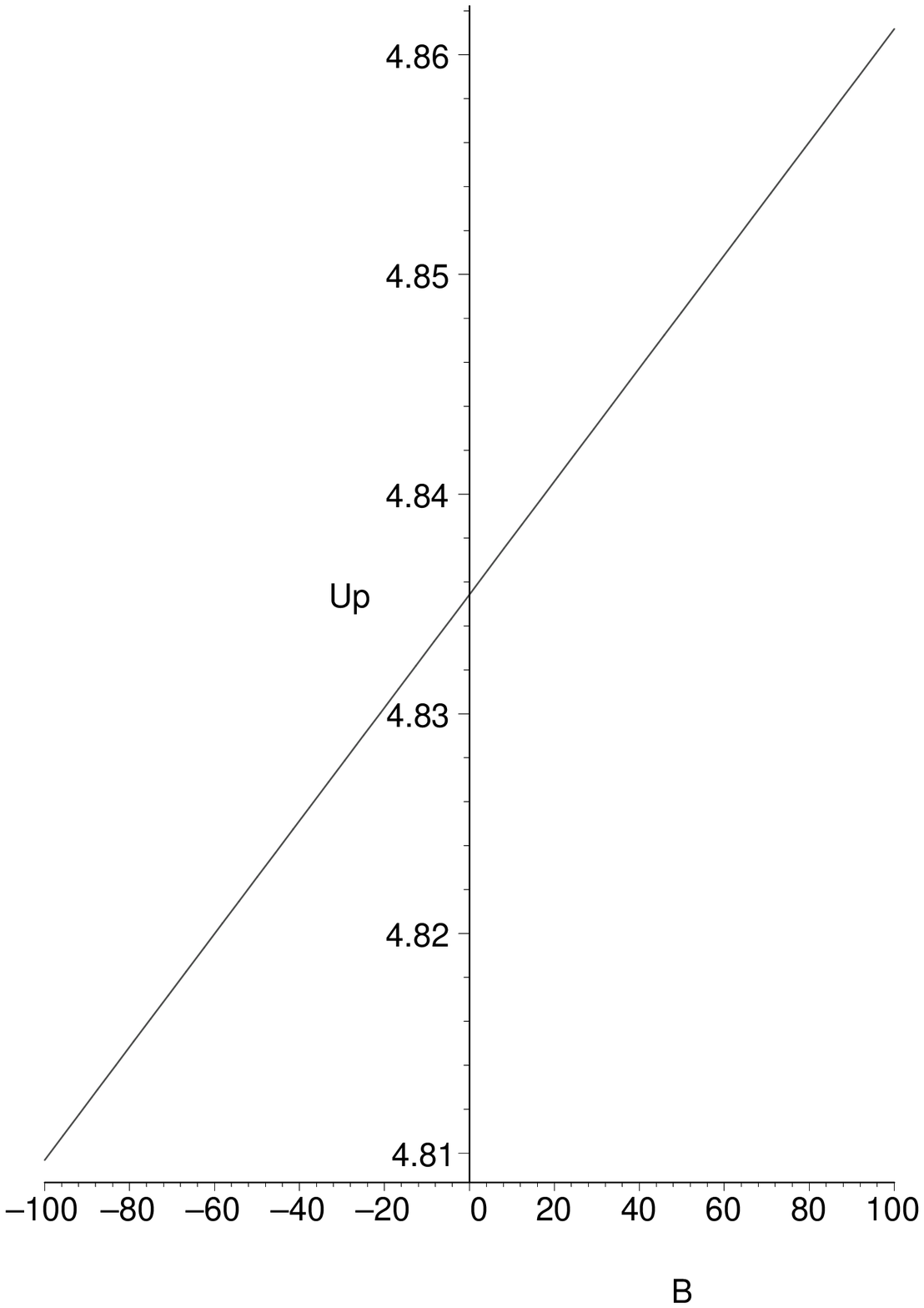,width=5.8cm,clip=}}
\caption{\small{
The scalar potential, after the axionic scalar extremization with 
the two possible signs $-$ (Um$\times10^{-91}$) and + (Up$\times10^{-91}$), 
at a fixed value of the orbifold
length ${\cal L}=10.82$ near the minimum, 
as a function of $B=b_2\chi$.}}
\label{} 
\end{figure}

\section{Summary and outlook}

In the present paper we have studied various implications of the eleven-dimensional $R^4$ term for heterotic M-theory. Working to order $\kappa^{4/3}$, we derived the change in the K\"{a}hler potential of the universal moduli and the ensuing corrections to the four-dimensional soft supersymmetry breaking terms. 
In particular, we observed that the induced $R^4$ corrections do not spoil the universality of the soft terms.

Next, we performed a detailed analysis of warp-factor geometric deformations
of the background metric in the presence of the $R^4$ term, and commented on the
generic-type deformations. 
We used the Killing spinor equations to extract the modified differential equations
obeyed by the warp factors. We have presented a few explicit solutions, depending on the nature of the background fluxes. This has enabled us to find the correction to
the volume of the Calabi-Yau three-fold, which turned out to be proportional to the
Calabi-Yau Euler number. 
Our careful averaging over the $x^{11}$ direction also showed that the positive definitness of the Calabi-Yau volume in the solution of \cite{CK} is a coordinate dependent statement. Thus, although we have found the expected shift of the volume of the Calabi-Yau manifold, it is unlikely that this can serve as a singularity resolution mechanism as suggested in \cite{CK2}.
As shown in Subsection 4.1, when neglecting higher derivative corrections and considering the warp-factor metric deformations to all orders in $\kappa$, one 
cannot turn on simultaneously the two flux components: the boundary flux $\beta_a$, which does not introduce any $x^{11}$ dependence in the solution, and the flux $\alpha$ which induces an $x^{11}$ fibration of the Calabi-Yau. 
However, taking into account higher derivative terms like $R^4$, $D^4R^4$ etc. 
induces changes beyond the order $\kappa^{4/3}$ to the 
supersymmetry variations and field equations. 
And so in principle this opens up the possibility to find solutions to higher orders which have both $\alpha\neq 0$ and $\beta_a \neq 0$. This would be of great interest, since both components are important: the $\alpha$-flux for stabilization of the orbifold length, while the $\beta_a$ flux for generating a superpotential that 
would allow the stabilization of the complex structure moduli of the 
Calabi-Yau. 
At the same time, it is likely that having turned on both $\alpha,\beta_a$ fluxes
one has to abandon the warp-factor deformations, and consider 
generic non-K\"{a}hler deformations of the Calabi-Yau  background.

The understanding of compactifications on six-dimensional non-K\"{a}hler 
manifolds is necessary for the stabilization of the complex structure moduli.
 Therefore, one natural direction for future research would be finding explicit non-K\"{a}hler backgrounds for the weakly coupled heterotic $E_8\times E_8$ string with methods similar to those of \cite{nonK} for the $SO(32)$ case. 
Another direction would be to study their strong coupling limit by 
trying to solve the generalized Hitchin flow equations we derived in Appendix B 
together with the appropriate $G$-field equation and Bianchi identity.

We have also studied the effective scalar potential in the context of an expansion of the background to order $\kappa^{4/3}$. Since the no scale structure of the K\"{a}hler potential for the $T$-modulus is not violated by the $R^4$ term, the main effect is not due to the higher derivative correction, as was the case in string theory \cite{BB}, but rather to the expansion of the background. 
We have found that de Sitter vacua still exist, although with a much smaller 
cosmological constant. 
It is worth investigating whether even higher order corrections, like  
$D^4R^4$ compactified on the zeroth order
background or $R^4$ compactified on the deformed background etc., would violate the no scale structure of the K\"ahler potential, thus leading to qualitative changes 
along the lines of \cite{BB}.

It would also be interesting to see how far one can get in building the 
eleven-dimensional action of Horava-Witten theory at higher orders, with the approach of \cite{IM}. Lastly, one could also address the issue of how a truncation to order $\kappa^{4/3}$ affects the assisted inflation solution of heterotic M-theory \cite{BBK}.

\section*{Acknowledgements}
We would like to thank K. Dasgupta, P. de Medeiros, M. Haack, A. Krause and A. Tseytlin for useful conversations. This work was supported in part by DOE grant DE-FG02-95ER40899.

\appendix

\section{Derivation of the K\"{a}hler potential}
\setcounter{equation}{0}

The Christoffel symbols for the metric

\be
ds^2 = e^{-6a-c} \bar{g}_{\mu \nu} dx^{\mu} dx^{\nu} + e^{2a} \tilde{g}_{mn} dx^m dx^n + e^{2c} (dx^{11})^2
\ee
are
\bea
&&\Gamma^{\,\, \mu}_{n \,p} = - e^{8a+c} \tilde{g}_{np} \partial^{\mu} a \,\, , \qquad \Gamma^{\, m}_{n \,\rho} = \delta^m_n \partial_{\rho} a \,\, , \qquad \Gamma^{\,\,\,\, \mu}_{11 \,11} = - e^{6a+3c} \partial^{\mu} c \,\, , \nn \\ \nn \\
&&\Gamma^{\,\, 11}_{11 \, \mu} = \partial_{\mu} c \,\, , \qquad \Gamma^{\, \rho}_{\mu \nu} = \bar{\Gamma}^{\, \rho}_{\mu \nu} - \delta^{\rho}_{( \mu} \,\pd_{\nu )} (6a+c) + \frac{1}{2} \eta_{\mu \nu} \pd^{\rho} (6a+c) \,\, . \label{Chr_sym}
\eea
Note that also $\Gamma^{\,\,\, 11}_{11 \,11} \neq 0$, but it will not be needed.\footnote{All partial derivatives in (\ref{Chr_sym}) and below are w.r.t. the metric $\bar{g}_{\mu \nu}$, i.e. they do not include the warp factor: all dependence on the moduli $a$ and $c$ is written down explicitly.}

The nonzero curvature components are
\bea \label{curv}
R^{\mu}{}_{m \nu n} &=& - e^{8a+c} \tilde{g}_{mn} \left[ 7 \pd_{\nu} a \pd^{\mu} a + \pd_{\nu} \pd^{\mu} a + \frac{1}{2}(\pd_{\nu} c \pd^{\mu} a + \pd_{\nu} a \pd^{\mu} c) \right. \nn \\
&-& \left. \frac{1}{2} \delta^{\mu}_{\nu} \, \pd^{\rho} a \, \pd_{\rho} (6a+c) \right] \, , \nn \\
R^m{}_{\mu n \nu} &=& - \delta^m_n \left[ 7 \pd_{\mu} a \pd_{\nu} a + \pd_{\mu} \pd_{\nu} a + \pd_{( \mu} a \,\pd_{\nu )} c - \frac{1}{2} \eta_{\mu \nu} \pd_{\rho} a \pd^{\rho} (6a+c) \right] \, , \nn \\
R^k{}_{mnp} &=& \tilde{R}^k{}_{mnp} - e^{8a+c} \,\pd_{\rho} a \pd^{\rho} a \, [\tilde{g}_{pm} \delta^k_n - \tilde{g}_{mn} \delta^k_p] \, , \nn \\
R^{\mu}{}_{\nu \rho \sigma} &=& \bar{R}^{\mu}{}_{\nu \rho \sigma} + ... \, , \nn \\
R^{11}{}_{\mu 11 \nu} &=& -2\pd_{\mu} c \pd_{\nu} c - \pd_{\mu} \pd_{\nu} c - 6 \pd_{( \mu} c \,\pd_{\nu )} a + \frac{1}{2} \eta_{\mu \nu} \pd_{\rho} c \pd^{\rho} (6a+c) \, , \nn \\
R^{\mu}{}_{11 \nu 11} &=& e^{6a+3c} \left[-2 \pd_{\nu} c \pd^{\mu} c - \pd_{\nu} \pd^{\mu} c - 3 (\pd_{\nu} c \pd^{\mu} a + \pd^{\mu} c \pd_{\nu} a) \right. \nn \\
&+& \left. \frac{1}{2} \delta^{\mu}_{\nu} \pd^{\rho} c \pd_{\rho} (6a+c)\right] \, , \nn \\
R^{11}{}_{m 11 n} &=& -e^{8a+c} \tilde{g}_{mn} \pd^{\rho} a \pd_{\rho} c \, , \nn \\
R^m{}_{11 n 11} &=& - e^{6a+3c} \delta^m_n \pd^{\rho} a \pd_{\rho} c \, . \label{Riem}
\eea
The expression $...$ in $R^{\mu}{}_{\nu \rho \sigma}$ is somewhat messy, so let us write down only what we need, namely its contribution to the scalar curvature:
\be
R^{\mu}{}_{\nu \rho \sigma} \delta^{\rho}_{\mu} e^{6a+c} \bar{g}^{\nu \sigma} = e^{6a+c} \left[ R^{(4)} + 3 (6 \pd_{\mu} \pd^{\mu} a + \pd_{\mu} \pd^{\mu} c) -\frac{3}{2} (6\pd a + \pd c)^2 \right] \, . \label{R_sc}
\ee

Now let us see how the action $S_0 + S_1$ in (\ref{S_0}), (\ref{S1}) reduces to an effective four-dimensional theory of gravity and the scalars $a,c$. Keeping only terms that are at most quadratic in derivatives and ignoring the flux contribution, we find from $S_0$:\footnote{For the reduction of the CJS action to fourdimensions, terms like $\Box a$, $\Box c$ can be dropped as they are simply total derivatives. But later on, when considering the $R^4$ correction, such terms will appear  multiplied by nontrivial functions and so will have to be kept.} 
\be
\int d^{11}x \sqrt{-g} R^{(11)} = Vol_7 \int d^{4}x \sqrt{-\bar{g}} \left[ R^{(4)} - 24 (\pd a)^2 - \frac{3}{2} (\pd c)^2 - 6 \pd a \pd c \right] \, , \label{S0}
\ee
where $Vol_7$ is the volume of the 6d space times the length of the interval; we have dropped the total derivative term $\Box (6a+c)$. Upon the field redefinition $\hat{c} = c + 2 a$ one recovers the action of \cite{LOW}.

Now let us consider the contribution of $S_1$. First, let us look at the following part of the integrand in (\ref{S_1}):
\be
\sqrt{-g} \,2^6 (12Z - RS + 12 R_{IJ}S^{IJ}) \, .
\ee
We can evaluate its contribution to the kinetic terms of the four-dimensional scalars $a$ and $c$, using that as in \cite{BBHL} 
\be
Z = 24 (2 \pi)^3 e^{6a+c} (\pd a)^2 Q
\ee
and computing the Ricci tensor and scalar curvature from (\ref{Riem}), (\ref{R_sc}).\footnote{The term $R_{IJ}S^{IJ}$ contributes only when the indices $I,J$ range over the 6d manifold, i.e. $I,J=m,n$.} We obtain
\be
(2 \pi)^3 \sqrt{-\bar{g}\tilde{g}} \, Q \,2^6 (-5 \cdot 12^2 (\pd a)^2 + 18 (\pd c)^2 -12 R^{(4)}) \, . \label{S1_1}
\ee
To get to the last expression, we have partially integrated terms of the form $Q \Box a$, $Q \Box c$ using that $\pd_{\mu} Q = - 6 Q \pd_{\mu} a $.

Finally, let us turn to the remaining term in $S_1$, $-\frac{1}{2} E_8$. The only index contractions in $E_8$ which give terms at most quadratic in derivatives are:
\bea
E_8&=&\frac 1{3!}4
\bigg(3 \epsilon^{\mu_1\mu_2 \mu_3\mu_4} \epsilon_{\mu_1\mu_2\nu_3\nu_4}
R_{\mu_3\mu_4}{}^{\nu_3\nu_4} +4\epsilon^{\mu_1\mu_2 \mu_3\mu}
\epsilon_{\mu_1\mu_2\mu_3 \nu} R_{\mu 11}{}^{\nu 11} \bigg)\nonumber\\
&&\epsilon^{m_1 m_2 \dots m_6}\epsilon_{n_1 n_2\dots n_6}
R_{m_1 m_2}{}^{n_1 n_2}
R_{m_3 m_4}{}^{n_3 n_4}R_{m_5 m_6}{}^{n_5 n_6}
\nonumber\\
&=&(-2^4)(12 \cdot 2^3 \pi^3)Q\frac 1{3!}4\bigg(3 \cdot 4 R_{\mu\nu}{}^{\mu\nu}+ 4 \cdot 6 R_{\mu 11}{}^{\mu 11}
\bigg),
\eea
where we have used that 
\bea
&{}&\!\!\!\!\!\!\!\!\!\!\!\!\!\!\!\!\!\!
\epsilon^{m_1 m_2 \dots m_6}
\epsilon_{n_1 n_2\dots n_6}R_{m_1 m_2}{}^{n_1 n_2}
R_{m_3 m_4}{}^{n_3 n_4}R_{m_5 m_6}{}^{n_5 n_6}=\nonumber\\
&{}&\!\!\!\!\!\!\!\!\!\!\!\!\!\!\!\!\!\!\!
\!\!-2^4 
\bigg(R_{m_1 m_2}{}^{m_3 m_4}R_{m_3 m_4}{}^{m_5 m_6}R_{m_5 m_6}{}^{m_1 m_2}
-2R_{m_1}{}^{m_3}{}_{m_2}{}^{m_4}R_{m_3}{}^{m_5}{}_{m_4}{}^{m_6}
R_{m_5}{}^{m_1}{}_{m_6}{}^{m_2}\bigg)\nonumber\\
&=&-2^4 (12 \cdot2^3\pi^3) Q \, .
\eea
From (\ref{Riem}), (\ref{R_sc}) we see that
\be
R_{\mu \nu}{}^{\mu \nu} + 2 R_{\mu 11}{}^{\mu 11} = e^{6a+c} \left[ R^{(4)} - 54 (\pd a)^2 -\frac{3}{2} (\pd c)^2 -6 \pd a \pd c + 18 \Box a + \Box c \right]
\ee
and hence
\bea
-\frac{1}{2} \sqrt{-g} E_8 &=& 2^6 (12 \cdot (2 \pi)^3) \sqrt{-g} Q (R_{\mu \nu}{}^{\mu \nu} + 2 R_{\mu 11}{}^{\mu 11}) = \nn \\
&=& 2^6 (12 \cdot (2 \pi)^3) \sqrt{-\bar{g} \tilde{g}} Q \left[ R^{(4)} + 54 (\pd a)^2 - \frac{3}{2} (\pd c)^2 \right]\!,  \label{E_8}
\eea
where we have partially integrated the terms containing $Q \Box a$ and $Q \Box c$.

Now assembling (\ref{S0}), (\ref{S1_1}) and (\ref{E_8}) we find the action
\bea
S_0 + S_1 &=& -\frac{V l}{2 \kappa^2_{11}} \int d^4x \sqrt{-\bar{g}} \, \bigg\{ R^{(4)} - \left[24 - \frac{6 \,b_2 \,\kappa^{4/3} \chi e^{-6a}}{V}\right] (\pd a)^2 \nn \\
&-& \bigg. \frac{3}{2} (\pd c)^2 - 6 \pd a \pd c \bigg\} \, ,
\eea
where
\be \label{b2}
b_2 = 12 \cdot 2^6 \cdot (2 \pi)^3 b_1 T_2 \cdot (2 \kappa^{2/3}_{11})
\ee
and the factor of $e^{-6a}$ is due to the fact that $Q$ in (\ref{chi}) was defined w.r.t. the metric $e^{2a} \tilde{g}_{mn}$ whereas the volume element we were now left with was just $\sqrt{\tilde{g}}$. Note that $b_2$ is a constant independent of $\kappa_{11}$ since $T_2 = (2 \pi)^{2/3} (2 \kappa^2_{11})^{-1/3}$. For convenience, from now on we normalize $V\equiv\int d^6 x \sqrt{\tilde{g}} = 1$. Clearly we can diagonalize the kinetic terms with the same field redefinition as before, i.e. $\hat{c} = c + 2 a$. The result is:
\be \label{f_action}
S_0 + S_1 = -\frac{l}{2 \kappa^2_{11}} \int d^4x \sqrt{-\bar{g}} \, \bigg\{ R^{(4)} - \,6 \,[3 - b_2 \,\kappa^{4/3} \chi e^{-6a}] \,(\pd a)^2 - \frac{3}{2} (\pd \hat{c})^2 \bigg\} \, .
\ee

Now we are ready to read off the new K\"{a}hler potential. Recall that the fields $a$ and $c$ make up the real parts of two chiral superfields \cite{LOW} with bosonic components
\be
S = e^{6a} + i \sigma_S \, , \qquad T = e^{\hat{c}} + i\sigma_T \, ,
\ee
where the axionic scalars 
$\sigma_S$, $\sigma_T$ originate from the eleven-dimensional three-form field $C$. For convenience we will use from now on $S$, $T$ to denote the full superfields. The kinetic terms of the four-dimensional $N=1$ effective action for these fields are of the standard form
\be
2 K_{S \bar{S}} dS d\bar{S} + 2 K_{T \bar{T}} dT d\bar{T} \, .
\ee
Since the $(\pd \hat{c})^2$ term in (\ref{f_action}) does not receive any correction, the corresponding part of the K\"{a}hler potential is as before:
\be
K_{(T)} = - 3 \ln (T + \bar{T}) \, .
\ee
On the other hand, the new kinetic term for $a$ can be reproduced from the following K\"{a}hler potential:
\be \label{K_S}
K_{(S)} = - \ln (S + \bar{S}) - \frac{b_2 \kappa^{4/3} \chi}{6 (S + \bar{S})} \, .
\ee

We also record the derivatives of the zero-th order K\"ahler potential 
\be
K_0=-\ln[(T+\bar T)^3 (S+\bar S)]\equiv -\ln(2^4\cV_{OM}^3\cV)
\ee
which are needed in the evaluation of the scalar potential
\be
K_{0,S}=-\frac{1}{2\cV},~~~~K_{0,T}=-\frac{3}{2\cV_{OM}},~~~~
K_{0,S\bar S}=\frac{1}{4\cV^2},~~~~K_{0,T\bar T}=\frac{3}{4\cV_{OM}^2} \, ,
\ee
where we introduced the notation $S+\bar S=2\cV$, $T+\bar T=2\cV_{OM}$.

\section{Generalized Hitchin flow equations}
\setcounter{equation}{0}

Dall'Agata and Prezas studied the conditions for $N=1$ compactifications of M-theory on seven-dimensional manifolds with $SU(3)$ structure \cite{DP}. Recall that requiring $G_2$ holonomy is too restrictive, meaning that it leads to trivial warp factors and no fluxes \cite{BJ}. On the other hand $G_2$ structure, considered in \cite{KMT}, carries less information than the $SU(3)$ structure case.

Let us summarize the relevant results of \cite{DP}. The eleven-dimensional metric is of the form:
\be \label{metric}
ds_{11}^2 = e^{2\Delta} \eta_{\mu \nu} dx^{\mu} dx^{\nu} + ds_7^2 \, ,
\ee
where $\Delta$ depends only on the coordinates of the internal 7d space. Since we are interested in an internal metric which is a warped product of an interval and a six-dimensional non-K\"{a}hler manifold whose three torsion classes $W_3$, $W_4$, $W_5$ are non-vanishing\footnote{Recall that 6d non-K\"{a}hler manifolds are classified by five torsion classes $W_i$, $i=1,...,5$ and that in heterotic string compactifications supersymmetry requires $W_{1,2}=0$, whereas generically all three $W_3$, $W_4$, $W_5$ can be nonzero. For more details see \cite{CCDL}.}, we take for $ds_7^2$ the form given in Section 4.2 of \cite{DP}:
\be \label{mans}
ds_7^2 = e^{p\phi(y,t)} ds_6^2(y,t)+e^{2\phi(y,t)}dt^2 \, ,
\ee
where $p$ is a real number and $t$ parametrizes the interval $I$. The 4-form flux $G$ has decomposition as in footnote 10 in terms of the 1-form $v$, 2-form $J$ and three-form $\Psi$ that define an $SU(3)$ structure in $d=7$. In fact, it will be more convenient to use the 7d Hodge dual with decomposition (see (3.11) of \cite{DP}):\footnote{As recalled in footnote 10, conditions (3.13) of \cite{DP} imply that $c_1'= 0$, $c_2'= 0$, $V = 0$ in (3.11) there.}
\be
*_7 G = \frac{Q}{3} J\wedge v + v \wedge A - \frac{3}{2} J \wedge \sigma + S \, .
\ee

The supersymmetry conditions can be rewritten as equations relating $v$, $J$, $\Psi$ and the $G$-flux components $Q,A,\sigma,S$.\footnote{The notation in this appendix is the same as in \cite{DP} and should not be confused with the notation in the rest of the present paper.} Recall that the six-dimensional derivatives of the three $SU(3)$-structure defining forms determine the non-K\"{a}hler manifold whereas their $\pd_t$ derivatives determine its fibration along the seventh dimension. Let us now write down these generalized Hitchin flow equations for arbitrary $p$ (unlike the case $p=\frac{1}{2}$ in eqs. (4.37), (4.38) of \cite{DP}) so that all of the torsion classes $W_i$, $i=3,4,5$, are non-vanishing. Using
\be
v = e^{\phi} dt\, , \qquad J = e^{p \phi} \hat{J} \, , \qquad \Psi = e^{\frac{3}{2}p\phi} \hat{\Psi} \, ,
\ee
where $\hat{}$ denotes a six-dimensional quantity, and also (3.20), (3.21), (4.34) of \cite{DP} we find for the six-dimensional space:
\bea \label{dJ}
\hat{d} \hat{J} &=& -2 e^{-p\phi} S -\left(p-\frac{1}{2}\right) \hat{d} \phi \wedge \hat{J} \nn\\
\hat{d} \hat{\Psi} &=& -\frac{3}{2}(p-1) \,\hat{d} \phi \wedge \hat{\Psi}
\eea
and for the dependence on the interval:
\bea \label{parJ}
\partial_t \hat{J} &=& - p \dot{\phi} \hat{J} + \frac{2}{3} Q e^{\phi} \hat{J} - 2 e^{-(p-1) \phi} A \nn \\
\partial_t \hat{\Psi} &=& -\frac{3}{2} p \dot{\phi} \hat{\Psi} +Q e^{\phi} \hat{\Psi} \, .
\eea
Clearly $W_4$ vanishes if $p=\frac{1}{2}$. And also one can easily see that substituting $p=\frac{1}{2}$ in (\ref{dJ}) and (\ref{parJ}), gives exactly (4.37), (4.38) of \cite{DP}. From (\ref{dJ}) we can read off the torsion classes of the 6d manifold:
\be \label{W}
W_3 = -2 e^{-p\phi} S \, , \qquad W_4 = -\left(p-\frac{1}{2}\right) \hat{d} \phi \, , \qquad W_5 = -\frac{3}{2} (p-1) \hat{d} \phi \, .
\ee
If this space is to be a solution of the weakly coupled heterotic string\footnote{Strictly speaking, this is not necessary in order to have a solution of the eleven-dimensional theory. But  backgrounds that have this property are most easily interpreted as eleven-dimensional lifts of ten-dimensional heterotic string solutions.}, then supersymmetry requires $W_5 = - 2 W_4$ (see \cite{CCDL}) and so fixes $p = \frac{5}{7}$.

However, the ansatz (\ref{mans}) is not the most general one. Compatibility of the structure of the solutions in \cite{DP} and \cite{KMT} requires that $\tilde{\phi} = - 2 \Delta$, where $\tilde{\phi}$ is the warp factor in front of $dt^2$ and $\Delta$ is the one in front of the four-dimensional space $\eta_{\mu \nu} dx^{\mu} dx^{\nu}$. Note that this is the same condition as (\ref{kb}). On the other hand, there is no reason why the warp factor in front of $ds_6^2(y,t)$ would be related to any of the other two. So let us consider metric of the following  form:
\be
ds_{11}^2 = e^{2\Delta} \eta_{\mu \nu} dx^{\mu} dx^{\nu} + e^{p\phi(y,t)} ds_6^2(y,t)+e^{-4 \Delta}dt^2 \, ,
\ee
where $\Delta$ and $\phi$ are unrelated to each other. It is easy to check that instead of (4.34) and (4.35) of \cite{DP} we have now $\sigma = \hat{d} \Delta$, \,$\dot{\Delta} = -\frac{1}{3} Q e^{-2 \Delta}$, \,$\hat{d} Q = 2 Q \hat{d} \Delta$. Therefore using $v = e^{-2 \Delta} dt$, $J = e^{p \phi} \hat{J}$, $\Psi = e^{\frac{3}{2} p \phi} \hat{\Psi}$ and (3.20), (3.21) of \cite{DP}, we derive for the  generalizations of (\ref{dJ}) and (\ref{parJ}):
\bea \label{dJn}
\hat{d} \hat{J} &=& -2 e^{-p \phi} S -p\hat{d} \phi \wedge \hat{J} - \hat{d} \Delta \wedge \hat{J} \nn\\
\hat{d} \hat{\Psi} &=& -\frac{3}{2} p \hat{d} \phi \wedge \hat{\Psi} - 3 \hat{d} \Delta \wedge \hat{\Psi}
\eea
and
\bea
\partial_t \hat{J} &=& - p \dot{\phi} \hat{J} + e^{- 2\Delta} \left( \frac{2}{3} Q \hat{J} - 2 e^{-p \phi} A\right) \nn\\
\partial_t \hat{\Psi} &=& -\frac{3}{2} p \dot{\phi} \hat{\Psi} + e^{-2 \Delta} Q \hat{\Psi} \, .
\eea
From (\ref{dJn}) we read off the following torsion classes:
\be
W_3 = -2 e^{-p\phi} S \, , \qquad W_4 = -p \hat{d} \phi - \hat{d} \Delta \, , \qquad W_5 = -\frac{3}{2} p \hat{d} \phi -3 \hat{d} \Delta \, .
\ee
If one wants the 6d non-K\"{a}hler manifold to be a solution of heterotic strings, i.e. to satisfy $W_5 = -2 W_4$, then one must impose a linear relation between $\hat{d} \Delta$ and $\hat{d} \phi$. Taking $\hat{d} \Delta = k \hat{d} \phi$ for some constant $k$, we find that $k = - \frac{7}{10} p$. It is easy to see that this is consistent with our previous  considerations: for $p=\frac{5}{7}$ we recover the relation $\hat{d} \Delta = -\frac{1}{2} \hat{d} \phi$ valid for the ansatz (\ref{mans}) \cite{DP}.


\end{document}